\documentclass[sn-mathphys-num,iicol]{sn-jnl}% Math and Physical Sciences Numbered Reference Style 
\usepackage{graphicx}%
\usepackage{multirow}%
\usepackage{amsmath,amssymb,amsfonts}%
\usepackage{amsthm}%
\usepackage{mathrsfs}%
\usepackage[title]{appendix}%
\usepackage{xcolor}%
\usepackage{textcomp}%
\usepackage{manyfoot}%
\usepackage{booktabs}%
\usepackage{algorithm}%
\usepackage{algorithmicx}%
\usepackage{algpseudocode}%
\usepackage{listings}%

\usepackage{bm}

\newcommand{\hs}[1]{ {\color{red}{#1}} }
\newcommand{\ga}[1]{ {\color{blue}{#1}} }

\begin{document}

\title[Droplets sitting on thin elastic sheets]{Droplets sitting on thin elastic sheets: A study with the boundary element method}


\author[1]{\fnm{Salik} \sur{Sultan}}\email{sultan@tu-berlin.de}

\author[1]{\fnm{Josua} \sur{Grawitter}}

\author[1]{\fnm{Gon\c calo C.} \sur{Antunes}}\email{g.antunes@tu-berlin.de}

\author*[1]{\fnm{Holger} \sur{Stark}}\email{holger.stark@tu-berlin.de}

\affil[1]{\orgdiv{Division of Theoretical Physics, Institute of Physics and Astronomy}, \orgname{Technische Universit\"{a}t Berlin}, \orgaddress{\street{Hardenbergstr. 36}, \city{Berlin}, \postcode{10623}, \country{Germany}}}


\abstract{Elasto-capillarity of a droplet wetting an elastic sheet provides an interesting system, both for fundamental and
applied research. The droplet sinks into the sheet and assumes the shape of a lens. To determine the equilibrium shape in simulations,
%A crucial question is the equilibrium shape of such a droplet. To answer it,} 
we formulate a
%ga{extend earlier} 
boundary element method (BEM)
% \ga{approaches} 
%to simulate 
 %\ga{enable simulations of} droplets sitting on an elastic sheet.% 
 extending
 %\ga{
 our earlier approaches, and apply the BEM to three specific protocols for the boundary conditions of the sheet.
% \ga{We apply this method to study the equilibrium droplet shape for three different protocols: a sheet whose sides are clamped, one whose sides are pulled isotropically, and one whose sides are pulled uniaxially. In all three cases, } %
%On
For a clamped elastic sheet, 
%the droplet sinks into the sheet and assumes a lens shape. Using 
we use various morphological metrics to demonstrate that the lens shape crucially depends on the sheet thickness.
Stretching the sheet isotropically, allows for an additional control parameter to influence the droplet shape and the tension in the sheet,
which we quantify by radial profiles of the azimuthal and radial elastic stresses. We further demonstrate how the focal length 
of a liquid lens can be tuned by varying the applied tension.
Finally, stretching the sheet along one direction, elongates the droplet, and the sheet shows folds and dimples. 
 %
% \ga{The introduction of an applied isotropic tension allows for an additional control parameter for the droplet shape and the tension in the 
%sheet,} which we quantify by radial profiles of the azimuthal and radial elastic stresses. \ga{We further demonstrate how the focal lens of a 
%liquid lens can be tuned by varying the applied tension.} \ga{In the case of a uniaxial stretch, the droplet is elongated,} and the sheet shows 
%folds and dimples. 
% which crucially depends on the sheet thickness, as we demonstrate with various morphological metrics. 
% Stretching the sheet isotropically, allows to control the tension in the sheet, which we quantify by radial profiles of the azimuthal 
% and radial elastic stresses. This means, when considering the droplet as a liquid lens, one can tune its focal length by tension.
%Finally, stretching the sheet uniaxially, elongates the droplet, and the sheet shows folds and dimples. 
%This is strongest for free lateral edges of the sheet.
%\gaa{I kept the gist of the abstract, just massaged the text a little bit.}
}

%\keywords{NN}



\maketitle

\section{Introduction}
\label{sec:intro}
Many biological and synthetic solids, including tissue, plastics, slender rods, and fibrous media
%arrays of slender rods, 
are mechanically flexible and respond differently to wetting by a liquid than rigid substrates~\cite{Duprat16,Bico18}. This phenomenon is 
termed elasto-capillarity. The flexibility of the substrates can vary depending on the material and environmental conditions. Furthermore, the 
direction-dependent flexibility of anisotropic materials gives rise to additional complexity, also regarding the statics and dynamics of 
the contact line~\cite{Andreotti20}. A particularly interesting property
%example 
is the Shuttleworth effect, %the strain-dependent surface stress
where the surface stress depends on the induced strain
\cite{Shuttleworth50,Mueller04,Andreotti16}.
%
%General Intro: Reformulate\\
%Shuttleworth effect: Bring it?\\ 
%To describe soft substrates, one needs to carefully distinguish between surface tension $\gamma$ and surface stress  $\boldsymbol{\Upsilon}$.
%While they are identical for liquid-vapour and liquid-gas interfaces, they obey Shuttleworth's relation for interfaces between liquids and 
%soft solids or sheets. We give it here
%\hs{Shuttleworth's relation??}
 %in Eulerian coordinates~\cite{Shuttleworth50,Mueller04,Andreotti16},
%\begin{equation}
%\boldsymbol{\Upsilon} = \gamma \boldsymbol{1}
% + \frac{\mathsf{d} \gamma}{\mathsf{d} \boldsymbol{\epsilon}} \, ,
%\label{eq.shuttle}
%\end{equation}
%where $\boldsymbol{\epsilon}$ is the two-dimensional strain tensor  of the substrate or sheet.
%The relation implies that, at the three-phase contact line, each interface exerts a force according to its surface stress $\boldsymbol{\Upsilon}$ 
%and the substrate exerts an additional force due to its internal stresses.
%\\
%\\
Thus, the competition of capillary and elastic stresses has been subject of intense research efforts in recent years~\cite{Mueller04,Andreotti16,Duprat16,Chen18,Bico18}. Particular emphasis was given to the region around the contact line, which
was visualized in experiments\
%\ga{can be observed experimentally} 
\cite{Park14,Xu17,Gorcum18}.
%
%In particular, the contact line region is characterized by three contact angle; the previously established liquid contact angle and its vapour and %substrate counterparts, which add up to $360^\circ$, \emph{i.e.}, the unit circle.
%How the unit circle is distributed among the phases was visualized in experiments, for example, in 
%Refs.~\cite{Park14,Xu17,Gorcum18}.
%

Droplets also show durotaxis, where they move towards softer regions of the substrate~\cite{Style13},
in contrast to living tissues~\cite{Alert19},
and even perform  stick-slip motion \cite{Gorcum18}. Experiments with synthetic hydrogels
demonstrate that the stiffness of the substrate can be %manipulated
tuned by light\ \cite{Lee18}. In theory,
%\ga{Theoretical efforts are also underway:} 
a unified numerical model for a droplet 
moving on a soft solid has been proposed recently~\cite{Aland21}, and a thin-film model reproduced the durotaxis of a
droplet\ \cite{Gomez20}.

%
%Durotaxis:\\
%Interestingly, droplets on a substrate with varying stiffness perform \emph{durotaxis}, \emph{i.e.}, they move down a stiffness gradient 
%toward softer regions of the substrate~\cite{Style13}. 
%Footnote: Notably, living tissue performs durotaxis in the opposite direction~\cite{Alert19}.
%The reason is that a softer compared to a more rigid substrate allows the droplet to reduce its total surface and thus take on a shape which 
%is closer to the minimum-energy shape of a sphere.
%Along its path, a droplet may undergo stick-slip motion if it moves above a critical speed~\cite{Gorcum18}.
%Furthermore, the stiffness of solids can be manipulated with light, as demonstrated in the case of synthetic hydrogels~\cite{Lee18}, 
%and thus durotaxis can be generated.
%For the case of a droplet moving on a soft solid, a unified numerical model has been proposed recently~\cite{Aland21}
%and a simple thin-film model of a droplet on a sheet reproduced durotaxis~\cite{Gomez20}.
%

A particularly interesting realization of a soft substrate are thin elastic sheets. Early experiments demonstrated appealing wrinkling patterns around 
%the sessile droplet 
sessile droplets \cite{Huang07,Toga13,Schroll13}. This observation sparked further experiments with highly deformable elastomeric 
sheets\ \cite{Schulman15}, where %under uniaxial stress 
the droplet shape becomes anisotropic under uniaxial stress\ \cite{Schulman17,Smith21}.
Droplets covered by an elastic sheet show atypical equilibrium shapes\ \cite{Schulman18}, and  %after an impact with high speed
%\ga{, including the case where 
they can become completely wrapped after a high-speed impact on the sheet.\ \cite{Kumar18}. 
Experiments with thin sheets under tension have been reviewed in Ref.\ \cite{Davidovitch18}
and detailed theoretical studies
%\ga{(albeit} 
%mostly analytical, 
%\ga{)} 
exist \cite{Shanahan85,Shanahan87,Davidovitch18,Kozyreff23,Nair23}, including an investigation of 
the motion of thin drops on an elastic sheet\ \cite{Li25}. %Finally, an appealing material system for fabricating elastic sheets is nanoporous 
%silicon\ \cite{Brinker20,Brinker21}.
Finally, we point to nanoporous silicon as a novel material,
%out the exciting possibilities offered by novel materials such as nanoporous silicon\ 
which can be used to fabricate actuatable elastic sheets \cite{Brinker20,Brinker21}.

Beyond exhibiting interesting phenomenology,  elastocapillary effects are opening new tech\-no\-lo\-gi\-cal avenues. For example, a droplet 
sitting on an elastic sheet deforms the sheet and assumes the shape 
of a lens\ \cite{Davidovitch18}. Micro-optofluidic lenses have been an attractive development since their shape and thereby their properties 
can easily be tuned by electric fields\ \cite{Berge00,Nguyen10,Chen21}, pressure used in commercially available lenses\ \cite{Fogle20}, 
as well as 
%by 
an applied tension, as we demonstrate explicitly in this article.
%also using electrowetting, which directly influences the contact angle~\cite{Berge00}.
Furthermore, a very topical field currently are biocondensates. %in the cell that 
They are present in the cell and wet the %cell
cell membrane \cite{Agudo21,Kusumaatmaja21,Wang24}. %,a system which 
This system has recently been modeled by a phase-field approach  \cite{Mokbel24}.

% on membranes:\\
%wetting of cell membranes by condensates in the cell: \cite{Agudo21}\\
%phase-field modelling of biocondensates on membranes \cite{Mokbel24}\\
%intercellular wetting \cite{Kusumaatmaja21}\\
%Biomolecular condensates mediate bending and scission of endosome membranes \cite{Wang24}

In this article we formulate a boundary element method (BEM) to simulate droplets sitting on an elastic sheet. The method follows 
our earlier approaches, where we studied droplets sitting on spatio-temporal wettability patterns and on undulating 
substrates\ \cite{Grawitter21,Grawitter21b,Grawitter24}. After extending the formalized BEM of Ref.\ \cite{Grawitter24} to include
elastic deformations of the sheet, we demonstrate its applicability. The lens shape of the droplet sitting on a clamped elastic sheet crucially
depends on the sheet thickness, which determines the in-plane elastic and bending moduli of the sheet. We also demonstrate that 
the microscopic contact angle, revealed by a close-up of the region around the contact line, agrees with Young's angle.
However, it differs from a macroscopic contact angle determined by the overall shapes of the two droplet halves. Stretching the sheet 
isotropically, allows to control the tension in the sheet, which we quantify by radial profiles of the azimuthal and radial elastic stresses. 
Furthermore, we argue that changing the lens shape under tension,  provides a means to tune the focal length of liquid optical lenses. 
Finally, when stretching the sheet uniaxially, the droplet elongates and the sheet shows folds and dimples. They are strongest for free 
lateral edges of the sheet.

The article is organized as follows. We begin by introducing our extended boundary element method in Section~\ref{sec:syssim},
report the results of the three specific protocols  in Section\ \ref{sec.results}, and summarize as well as conclude in Section~\ref{sec.concl}.

\section{%Reformulated 
Extended boundary element method}
\label{sec:syssim}

A droplet sitting on a thin elastic sheet deforms the latter due to the Laplace pressure that acts within the droplet and pushes
against the sheet. In this section, we establish the relevant dynamic equations to describe and simulate the relaxation towards 
the equilibrium lens shape of the droplet.

Central to the droplet dynamics %of the droplet
is the fluid flow within and at the surface of the droplet,
where the surface velocity governs changes in droplet shape.
%the droplet-shape changes. 
%\ga{the latter of which governs the change in droplet shape}. 
Fluid flow at low Reynolds number, as considered in this work, 
%are
is described by the Stokes equations, $\eta \bm{\nabla}^2 \bm{v} -\bm{\nabla}p=0$, and the incompressibility condition
$\bm{\nabla} \cdot \bm{v} = 0$. Here, $\eta$ is the shear viscosity and $p$ is pressure. The boundary integral equation solves the Stokes
equations by first determining the velocity field at the surface, from which the bulk velocity field can then be deduced \cite{Pozrikidis92,Kim05}.
%\ga{which is sufficient to then determine the bulk velocity field \cite{Pozrikidis92,Kim05}} 
This procedure is the basis of the boundary element method (BEM) \cite{Pozrikidis92}, which we implemented in 
Refs.\ \cite{Grawitter21,Grawitter21b} for droplets moving on a substrate with spatiotemporal variations of the wettability.

In a following article,
%\ga{Expanding on this work,} 
we considered a droplet on an undulating substrate and realized that we needed to reformulate the BEM for the droplet motion in a 
more formalized way\
%\ga{ using an enhanced BEM whose equations were obtained with a higher degree of rigor}
\cite{Grawitter24}. Essentially, 
%\ga{Succinctly,} 
we developed equations for the overdamped dynamics of the droplet surface, 
where part of the friction matrix
% \ga{(relating velocities to surface tractions) }
is derived from the boundary integral equation and the driving force is the derivative of the 
droplet free energy. Here, we follow the same route %and strongly rely on but extend our earlier method. 
but need to extend this method. 
%\ga{GA: this makes it seem that we are using Grawitter24, when Salik uses Grawitter21, no?} 
%\hs{HS: Salik uses Grawitter24}
Besides the surface velocity field of the droplet, $\dot{\bm{R}}_0$, we also introduce the velocity field $\dot{\bm{R}}_{\text{sh}}$ of the two-dimensional elastic sheet, where
$\bm{R}_0$ and $\bm{R}_{\text{sh}}$ describe the shape of the droplet surface and elastic sheet, respectively. The dynamic equations
become:
\begin{equation}
\bm{G} \dot{\bm{R}} =
\left[ \begin{array}{cc}
\bm{G}_0& \bm{G}_\text{slip} \\
\bm{G}_\text{slip}^\text{t}& \bm{G}_\text{sh}
\end{array}
\right]
\left[  \begin{array}{c} 
\dot{\bm{R}}_{0} \\ \dot{\bm{R}}_{\text{sh}}
\end{array}
\right]
= - \bm{\nabla}_{\bm{R}}  \mathcal{F} 
%= \left[  \begin{array}{c} 
%\bm{\nabla}_{\bm{R}_0} \\ \bm{\nabla}_{\bm{R}_\text{sh}}
%\end{array}
%\right]
%\mathcal{F} 
\, .
\label{eq.dynamic}
\end{equation}
Here, $\mathcal{F} = \mathcal{F}_0 + \mathcal{F}_\text{sh} + \mathcal{F}_\text{cons}$ is the total free energy, which consists
of the free energies of the droplet ($\mathcal{F}_0$) and the elastic sheet ($\mathcal{F}_\text{sh}$), while
%\ga{plus} 
$\mathcal{F}_\text{cons}$
% \ga{which encodes for} 
contains constraints. They enforce constant droplet volume, with
a pressure $p_0$ as Lagrange parameter, and keep part of 
the elastic sheet always in contact with the droplet.
%\ga{. These being a constant droplet volume (with the pressure $p_0$ as Lagrange parameter) \ga{GA; Is $p_0$ the full pressure? 
%If the droplet moves, the pressure should not be homogeneous.}, and ensured contact between a part of the elastic sheet and the droplet.}
% to keep the volume fixed, with the pressure $p_0$ as Lagrange parameter,  and to keep parts of the elastic sheet in contact with 
%the droplet. 
%\\ \hs{HS: Yes, $p_0$ is the full pressure.} \\
The force $-\bm{\nabla}_{\bm{R}_0} \mathcal{F}_0$ together with $p_0$ provides the Laplace pressure and 
Young's force at the contact line (see Ref.\ \cite{Grawitter24}), whereas $-\bm{\nabla}_{\bm{R}_\text{sh}} \mathcal{F}_\text{sh}$ contains 
in-plane elastic shear and dilatation
forces as well as bending forces. The friction matrix $\bm{G}_0$ is due to viscous shear flow as described by the boundary integral equation 
and contains contact-line friction, while $\bm{G}_\text{sh}$ introduces %some 
a phenomenological friction coefficient to dampen the sheet motion.
It %might also 
may also be interpreted as, e.g., %capture
capturing the effect of an additional fluid phase in contact with the other side of the elastic sheet.
% \ga{, potentially capturing the effect of an additional fluid phase in contact the underside of the elastic sheet GA: would this work?}.
Liquid-substrate friction between droplet and elastic sheet is quantified by $\bm{G}_\text{slip}$ and also contributes to $\bm{G}_0$, 
$\bm{G}_\text{sh}$. 

In Sections\ \ref{subsec.droplet_constraint} - \ref{subsec.final_dynamics} we describe all contributions to Eq.\ (\ref{eq.dynamic}), 
also based on Ref.\ \cite{Grawitter24},  where our approach is fully justified,
%\ga{compared to experiments,} 
and in Section\ \ref{subsec.para} we non-dimensionalize the equations, introduce relevant parameters, and summarize simulation details.

\begin{figure} %[ht!]
  \centering
  \includegraphics[width= 1\columnwidth]{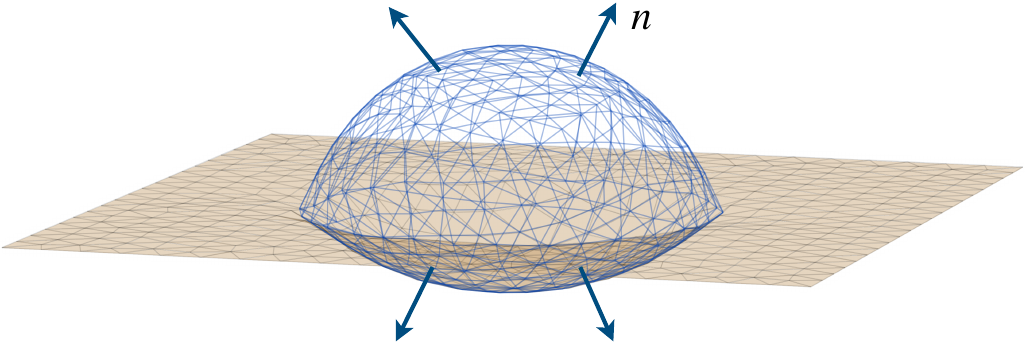}
\caption{Mesh of gridpoints for the droplet surface [blue, $\bm{R}_0 = (\bm{r}_1, \bm{r}_2, \bm{r}_3, \ldots )$] and the sheet
[brown, $\bm{R}_\text{sh} = (\bm{r}_\text{sh1}, \bm{r}_\text{sh2}, \bm{r}_\text{sh3}, \ldots)$]. The surface normals $\bm{n}$ point away from the
droplet volume.} 
\label{fig.mesh}
\end{figure}

As illustrated in Fig.\ \ref{fig.mesh} we approximate droplet surface and elastic sheet by two meshes so that configuration vector 
$\bm{R}_0 = (\bm{r}_1, \bm{r}_2, \bm{r}_3, \ldots )$ contains all vertices of the droplet mesh and $\bm{R}_\text{sh} = 
(\bm{r}_\text{sh1}, \bm{r}_\text{sh2}, \bm{r}_\text{sh3}, \ldots)$ all vertices of the sheet mesh. Then the force 
$- \bm{\nabla}_{\bm{R}}  \mathcal{F}$ becomes a high-dimensional column vector and $\bm{G}$ a matrix. 
%\\ \hs{HS: Yes, it is a very long vector.}

%\ga{GA: I dont understand what high dimensional vector means here. DO you want to say it is a very long vector? Or that all three dimensions fit in %this one vector because we pack values of x, y, and z in one column?}

%\hs{HS:} Remarks about boundary conditions, when we treat single interfaces/surfaces.

\subsection{Droplet-sheet free energy and constraints}
\label{subsec.droplet_constraint}

First, we collect and explain the different contributions to the free energy 
$\mathcal{F} = \mathcal{F}_0 + \mathcal{F}_\text{sh} + \mathcal{F}_\text{cons}$.

\subsubsection{Droplet free energy} 
%and constraints}
\label{subsubsec.droplet_free}

The droplet free energy is determined by integrating the surface tension over the surfaces of the droplet and the substrate, which in our case
is the elastic sheet. %For completeness, we 
%need to 
%add a term that contains a phenomenological constant for a line tension of the contact line.
%We motivate it further below but ultimately neglected it. \ga{GA: I don't think we need to mention that we neglect it here, it becomes confusing}
For the first part of the free energy, we rely on the form used in 
Ref.\ \cite{Grawitter24} and obtain in total
% we wrote the energy in the form
\begin{equation}
\mathcal{F}_0 = \gamma_\text{lg} A_\text{lg} - \gamma_\text{lg} \int\limits_{A_\text{sl}} \cos\theta_\text{eq}(\bm{r}) \text{d}^2\bm{r} 
+ \tau \int_\text{cl}  \sin \theta \text{d}r \, .
%\mathrm{const.} \, ,
\label{eq.F0}
\end{equation}
Here, $\gamma_\text{lg}$ is the surface tension and $A_\text{lg}$ the area of the liquid-gas interface. We also introduce 
the equilibrium contact angle by using Young's law,  $\gamma_\text{sl}-\gamma_\text{sg} = - \gamma_\text{lg} \cos\theta_\text{eq} $,
and the area $A_\text{sl}$ of the substrate-liquid interface. In what follows, we do not vary the contact angle, so 
$- \gamma_\text{lg} \cos\theta_\text{eq}$ just represents the difference of the constant surface tensions of the substrate-liquid and the substrate-gas
interfaces. Thus, we have 
\begin{equation}
\mathcal{F}_0 = \gamma_\text{lg} A_\text{lg} - \gamma_\text{lg} \cos\theta_\text{eq} A_\text{sl} + \tau \int_\text{cl}  \sin\theta \text{d}r \, .
\end{equation}
Note that, strictly speaking, this formula neglects changes in the total sheet area during deformation.

To motivate the last term on the right-hand side, we note that at the contact line the liquid-gas surface tension pulls up a ridge from a soft 
substrate with a lateral extension given by the elastocapillary length $l_\text{el} = \gamma_\text{lg} / E$, where $E$ is the Young modulus 
of the sheet material \cite{Andreotti16,Andreotti20}.  As we show in Section\ \ref{subsec.para}, this length is of the order of  
$0.06 \mathrm{\mu m} $, much smaller than the mesh size in our simulations.
%\ga{droplet radius, the length scales associated to the deformation of the sheet, and what one can feasibly take as the mesh size in our simulations. }  
In such a case, where this deformation is not resolved, 
%We do not resolve this deformation,  
one can account for the respective deformation energy by an effective line tension $\tau$.
%$\tau \propto \gamma_\text{lg}$. 
However, for elastic sheets this line tension can be safely neglected as stated in Ref.\ \cite{Andreotti20}.
Hence, 
%\ga{As a result,} 
we performed all our simulations with $\tau = 0$.

%, which scales like $\gamma_\text{lg}$ and which 
%we integrate along the contact line. The factor $\sin \theta$ with the dynamic contact angle $\theta$ comes in, since  for $\theta = 90^\circ$ 
%the largest amount of material is pulled up.
%%\\ \hs{HS: How do you determine again $\theta$?}
%%\\ \hs{HS: I think $\tau \propto \gamma_\text{lg}$}

\subsubsection{Constraints}
\label{subsec.cons}

There are two constraints in the droplet-sheet system, which we add to the free enery using Lagrange multipliers. First, the volume
of the droplet is constant and, second, the droplet and sheet surfaces are tightly connected to each other at the common interface.
%\ga{. Secondly, the droplet and sheet surfaces are tightly connected to each other at the common interface.} 
Both constraints contribute the free energy
\begin{equation}
\mathcal{F}_\text{cons} = p_0(V_0-V)+ \int\limits_{A_\mathrm{sl}} \lambda(\bm{x}) \left[z - z_\text{sh}(\bm{x})\right] \text{d}^2 \bm{r} \, ,
\label{eq.Fcons}
\end{equation}
%\begin{equation}
%\mathcal{F}_\text{cons} = p_0(V_0-V)+ \int\limits_{A_\mathrm{sl}} \lambda(\bm{x}) \left[z - h(\bm{x})\right] \text{d}^2 \bm{r} \, ,
%\label{eq.Fcons}
%\end{equation}
where the first term on the right-hand side constrains the droplet volume to the value $V_0$ and the Lagrange multiplier $p_0$ is %the 
a pressure. 
%\ga{GA: Same question as before here. What pressure is this?}. 
For computational purposes we note that with the help of Gauss's theorem the droplet volume can be expressed as 
$V=\frac{1}{3}\oint \bm{r} \cdot \bm{n} \text{d}^2\bm{r} $, where the unit vector $\bm{n}$ points along the outward normal. Furthermore, 
we use the Monge form to parametrize how the flat elastic sheet in the $x$-$y$ plane is deformed by the droplet along the vertical $z$ axis 
using the height function, \emph{i.e.}, the vertical coordinate of the sheet, $z_\text{sh}(\bm{x})$
%$h(\bm{x})$ 
with $\bm{x} = (x,y)$. So, the second term constrains the height $z$ of the droplet-sheet interface 
to the height variations of the sheet using the Langrange multipliers $\lambda(\bm{x})$. 

For the implementation in a simulation code, it makes sense to write the discretized version of the constraints in Eq.\ (\ref{eq.Fcons})
in the compact form 
\begin{equation}
\mathcal{F}_\text{cons} = \sum \limits_{i=0}^N \lambda_i g_i(\bm{R}) \, ,
\label{eq.Fcons2}
\end{equation}
where we set $\lambda_0 = p_0$ and $g_0 = V_0-V$. Furthermore, the droplet-sheet interface is discretized by $N$ nodes and
we have for the connectivity constraint of the $i$-th node,  $g_i(\bm{R}) =  z_i - z_{\text{sh}i}(x_i, y_i) $, 
%$g_i(\bm{R}) =  z_i - h(x_i, y_i) $
with $i =1 ... N$, while an area element is subsumed into $\lambda_i$.
%
%, $\sum_{i=1}^{N} \lambda_i g_i(\bm{R}) = \sum_{i=1}^{N} \lambda_i [z_i - h(x_i, y_i)] $.
%

%\hs{HS:
%1. \textbf{Salik}, I now use $z_{\text{sh}i}(x_i, y_i)$ instead of $h$. Please check.
%%2. Shall we add the droplet force Eq. (11) from Josuas paper , just to show what the physics is. We then also need to 
%%add the curvature force from the contact line. Is this all?
%}

\subsubsection{Sheet free energy}
\label{subsubsec.sheet_energy}

The sheet free energy consists of two contributions that describe in-plane shear and dilatation deformations, as well as
out-of-plane bending,
$\mathcal{F}_\text{sh} = \mathcal{F}_\text{d} + \mathcal{F}_\text{b}$.
% \ga{with $\mathcal{F}_\text{d}$ and $\mathcal{F}_\text{b}$.
%% capturing in- and out-of-plane deformations, respectively.}. 
We have used both free energies in previous work\ \cite{Schaaf17,Patel21}.

\textbf{Skalak model}
To consider in-plane deformations, we rely on the Skalak model developed to address deformations of red-blood cells as one example
of a hyperelastic material\ \cite{Skalak73}. 
%
% reference Skalak model
%
We describe any material point of the undeformed plane sheet in the $x$-$y$ plane by the 
two-dimensional vector $\bm{\xi} = (\xi_1, \xi_2)$, which due to in-plane and bending deformations moves to a point with the 
three-dimensional position vector $\bm{r}_\text{sh}$. With the help of the deformation gradient tensor 
$F_{i\alpha} = \partial r_{\text{sh},i}/ \partial \xi_\alpha$ ($i=1,2,3$ and $\alpha = 1,2$), we introduce the Cauchy strain tensor 
$\bm{C} =  \bm{F}^\text{t} \bm{F}$ with components $C_{\alpha\beta} = F_{i\alpha} F_{i\beta}$ that completely describes in-plane shear and 
dilatation deformations. The components $C_{\alpha\alpha}$ and $C_{12}$ quantify, respectively, stretching factors along the
$\alpha$-directions and the sheared angle between directions 1 and 2. 

The Skalak model uses invariants of the Cauchy strain tensor, $I_1 = \text{tr} \bm{C} - 2$ and $I_2 = \text{det} \bm{C} -1$, to construct
an elastic free energy also valid for large deformations,
\begin{eqnarray}
%\begin{equation}
  \mathcal{F}_\text{d} & = & \int  f_\text{d}  \,  \text{d}^2\xi    \label{eq.Skalak} \\  
  & = & \int  \Big[\frac{\kappa_S}{12} (I_1^2 + 2 I_1 - 2 I_2) + \frac{\kappa_A}{12} I_2^2\Big] \,  \text{d}^2\xi \, ,
\nonumber
%
%  \mathcal{F}_\text{d} = \int  \Big[\frac{\kappa_S}{12} (I_1^2 + 2 I_1 - 2 I_2) + \frac{\kappa_A}{12} I_2^2\Big] \,  \text{d}^2\xi \, ,
%
\end{eqnarray}
%\end{equation}
{where $f_d$ is the surface free energy density, and} $\kappa_S$ and $\kappa_A$ are the respective shear and area dilatation moduli. Note that the integral is performed over the
material coordinates $\xi_1$ and $\xi_2$. Now, when calculating the elastic forces $-\bm{\nabla}_{\bm{R}_\text{sh}} \mathcal{F}_\text{d}$
in the discretized form for the dynamic equation (\ref{eq.dynamic}), we rely on a procedure outlined in Ref.\ \cite{Krueger12} that explicitly 
considers deformations of the single mesh triangles.

For completeness, for small deformations 
we connect the Skalak model to the theory of linear elasticity,
% \ga{, which is valid for small deformations.}%, 
where the Cauchy strain tensor
%\ga{The Cauchy stress tensor} 
becomes $\bm{C}= 1 + 2 \bm{\epsilon}$ 
%$\bm{\sigma}= 1 + 2 \bm{\epsilon}$ 
and $\bm{\epsilon}$ is the conventional strain tensor. 
%\ga{GA:We already used $\bm{C}$ for the Cauchy strain tensor. Plus, we use $\bm{\sigma}$ to refer to the cauchy stress tensor 
%inside the droplet.}  
Then the Skalak free energy can be written in the form,
\begin{equation}
  \mathcal{F}_\text{d} = \int \Big[ \mu_\text{2D} \bm{\epsilon} \cdot \bm{\epsilon} + \frac{\lambda_\text{2D}}{2} (\text{tr} \bm{\epsilon})^2 \Big] \, 
  \text{d}^2r_\text{sh}
\end{equation}
%\ga{GA: Why did we switch from $\xi$ to $r_{sh}$?}
with $\bm{\epsilon} \cdot \bm{\epsilon} = \epsilon_{ij}\epsilon_{ij}$ and where the Lam\'e constants in two dimensions become 
$\mu_\text{2D}= \kappa_S/3$ for pure shear deformations 
%(\ga{GA: Why "for shear deformations?" and why does this qualifier also not apply to $\lambda_{2D}$?)} 
and $\lambda_\text{2D} = 2 \kappa_A/3$. The bulk modulus for pure area 
dilatations ($\bm{\epsilon} \propto \bm{1}$) reads $K_A = \mu_\text{2D} + \lambda_\text{2D} = (\kappa_S + 2 \kappa_A)/3$.
%\\ \hs{HS: \textbf{Salik}, are the formulas correct?}
%\\ \hs{HS: 1. The linear theory does not need to distinguish between material and spatial coordinates, typically one uses the spatial coordinates.}

\textbf{Bending free energy}
The bending or Helfrich free energy of the elastic sheet solely depends on the mean curvature, $1/R = (1/R_1 + 1/R_2) /2$, when the 
topology of the sheet stays the same \cite{Helfrich73}.
%\ga{, with 
Here, $R_1$ and $R_2$ are the principal radii of curvature. 
For a sheet discretized by a mesh of triangles, it can be shown that $\mathcal{F}_\text{b}$
depends on the relative orientation of neighboring triangles \cite{Kantor87,Gompper96}. Thus, we have
\begin{equation}
\mathcal{F}_\text{b} =  \frac{\kappa_b}{2} \int \Big( \frac{1}{R} \Big)^2 \text{d}^2r_\text{sh} \,\, \longrightarrow \,\,
%\mathcal{F}_\text{b} = 
\frac{\sqrt{3}\kappa_b}{2} \sum_{\langle i,j \rangle} \theta_{ij} \, ,
\end{equation}
where $\kappa_b$ is the bending constant. Furthermore, $\theta_{ij} = \arccos (\bm{n}_i \cdot \bm{n}_j)$ is the angle between 
the normal unit vectors $\bm{n}_i$, $\bm{n}_j$ of two neighboring triangles and the sum runs over all pairs of neighboring triangles.
From this form of $\mathcal{F}_\text{b}$, the bending force $-\bm{\nabla}_{\bm{R}_\text{sh}} \mathcal{F}_\text{b}$ can be readily calculated.

%Introduce formalized constraints .... ?????

\subsection{Friction matrix }
\label{subsec.friction}

In this section we address the friction matrix $\bm{G}$ of Eq.\ (\ref{eq.dynamic}) with its three contributions.
This includes friction of the droplet motion due to viscous shear stresses in the fluid flow and due to the motion of the contact line,
some phenomenological friction of the sheet to dampen its motion, and liquid-substrate friction between droplet and elastic sheet.

We first introduce the different contributions in Sections\ \ref{subsubsec.shear} - \ref{subsubsec.liquid-substrate} and then show
the total friction matrix in Section\ \ref{subsubsection.total}.

\subsubsection{Shear and contact-line friction}
\label{subsubsec.shear}

%\ga{We first deal with the friction induced by dissipation in the droplet $\bm{G}_0$ which we split in two contributions
%\begin{equation}
%\bm{G}_0 = \bm{G}_0^{(1)} + \bm{G}_0^{(2)},
%\end{equation}
%where $\bm{G}_0^{(1)}$ handles viscous dissipation in the bulk and near the contact line, and $ \bm{G}_0^{(2)}$ handles friction with the elastic %sheet. We now describe the former contribution.}

In Ref.\ \cite{Grawitter24} we explain in detail how shear and contact-line friction
%the two contributions 
are 
%\ga{$\bm{G}_0^{(1)}$} is 
derived, respectively,  from the boundary-integral equation and from an expression that relates viscous shear stresses %due to
to the contact-line motion. We do not repeat the derivation but mainly 
explain the final expressions. The friction matrix of the fluid droplet for the two contributions 
%\ga{the shear and contact-line friction} 
reads
\begin{equation}
 \bm{G}_0^{(1)} = \bm{X}^{-1}(\bm{ C}_d + \bm{Y}) +  \bm{P}_\text{cl}^\text{t}\bm{M} \bm{P}_\text{cl} \, .
 \label{eq.friction_droplet}
\end{equation}
% \ga{GA: We had already used $\bm{C}$ for the Cauchy strain tensor }
A further contribution $\bm{G}_0^{(2)}$ is due to liquid-substrate friction and addressed below.

%\\ \hs{HS: I went back to the old version but added an orienting sentence before 2.2.1}

%\ga{I think it was a bit confusing the way it was, I shuffled the explanation around a bit}

The first term on the right-hand side of Eq.\ (\ref{eq.friction_droplet}) derives directly from the boundary-integral equation. The components 
of the three matrices assume the following values:
\begin{align}
 X_{ij} & = \int_{C_j} \bm{O}(\bm{r}_i - \bm{r}) \, \text{d}^2r / A_j \\
 Y_{ij} & =\int_{C_j} \bm{T}_\text{O}(\bm{r}_i - \bm{r}) \bm{n} \, \text{d}^2 r \, ,
\end{align}
%\ga{GA: do we not have have $/A_j$ in the equation for $Y_{ij}$?}
%\hs{HS: No $A_j$ in $Y_{ij}$.}\\
where $\bm{O}$ is the Oseen tensor and $\bm{T}_\text{O}$ the associated stress tensor. The surface integral is performed over the polygonal 
cell $C_j$ with area $A_j$ constructed around mesh vertex $j$ and $\bm{r}_i$ is the position of vertex $i$. The matrix $\bm{C}_d$ is diagonal with
elements $c_i=1/2$, except if vertex $i$ is on the contact line, then $c_i=\theta_i / 2\pi$ with contact angle $\theta_i$.

The second term on the right-hand side of Eq.\ (\ref{eq.friction_droplet}) contains the friction matrix $\bm{M}$ of the contact line and follows from Moffatt's early result \cite{Moffatt64} that relates the normal 
stresses  along the liquid-gas interface to the contact-line velocity. After some integration\ \cite{Grawitter24}, one obtains
\begin{equation}
M_{ij}= 2\eta \frac{A_i}{\zeta}\ln\left(\frac{\zeta}{l_\mathrm{s}}\right) \frac{\sin^2 \theta_i}{\theta_i - \sin \theta_i \cos \theta_i} \delta_{ij} \, ,
\label{eq_cl_friction_matrix}
\end{equation}
where $\theta_i$ is the contact angle at vertex $i$, $l_s$ is the liquid-substrate slip length, which we introduce further below, and 
$\zeta$ is a mesoscopic length, which in practice acts as a fitting parameter. Finally, $\bm{P}_\text{cl}$ is a projection operator that projects 
the vertex velocities of the contact line out of the droplet-mesh velocity vector $\dot{\bm{R}}_0$ but only takes the velocity components 
normal to the contact line and tangential to the substrate.

\subsubsection{Elastic-sheet friction}

%\ga{We now introduce the friction owing to dissipation in the elastic sheet $\bm{G}_\text{sh}$, which we split into two contributions
%\begin{equation}
%\bm{G}_\text{sh}= \bm{G}_\text{sh}^{(1)} + \bm{G}_\text{sh}^{(2)},
%\end{equation}
%with the first contribution $\bm{G}_\text{sh}^{(1)}$ encoding for an effective dissipation whenever the sheet moves. To this end, we introduce a}

For the friction of the elastic sheet ,we introduce a phenomenological constant $\xi_\text{sh}$ mainly for computational reasons to 
dampen the dynamics of the elastic sheet and thereby ensure numerical stability, but also to capture dissipation inside the sheet and when there
is contact to a second fluid.
%\ga{with the dual purpose of capturing additional dissipation (e.g. a secondary fluid phase underneath the sheet or membrane viscosity), 
%and of ensuring numerical stability.}
The chosen value assures that the sheet relaxes much faster than the droplet. Thus, we write
%\\ \hs{HS: \textbf{Salik}, is the sentence correct?\\}
\begin{equation}
\bm{G}_\text{sh}^{(1)} = \xi_\text{sh} \bm{1} \, .
\label{eq.sheet_friction}
\end{equation}
An additional term $\bm{G}_\text{sh}^{(2)}$ is due to liquid-substrate friction, which we introduce now.
%\ga{The second term $\bm{G}_\text{sh}^{(2)}$ encodes the droplet-substrate friction, which we now introduce.}

\subsubsection{Liquid-substrate friction}
\label{subsubsec.liquid-substrate}

The friction between substrate and the liquid droplet is governed by the Navier condition, 
%$
\begin{equation}
l_s \bm{P}_\text{t} \bm{\sigma}\bm{n} + \eta \bm{P}_\text{t} (\bm{v} - \bm{v}_\text{sh}) = 0 \, .
%
%l_s \bm{P}_\text{t} \bm{\sigma}\bm{n} = \eta \bm{P}_\text{t} (\bm{v} - \bm{v}_\text{sh}) \, .
%
\label{eq.Navier}
\end{equation}
%$.
It is a pure kinematic condition 
%\ga{GA: I'm not sure what you mean with "pure kinematic" here. Isn't this equation a force balance between a driving force (left hand side) 
%and a friction force (right hand side)?} 
%\\ \hs{HS: $\bm{\sigma}$ is essentially the viscous stress tensor. So $\eta$ on both sides drops and the relative velocity is determined
%by velocity gradient and $l_s$. I was mentioning this a few times.} \\
and sets the relative velocity $\bm{v} - \bm{v}_\text{sh}$ with which the droplet slips along the substrate. 
Here, $\bm{P}_\text{t} \bm{\sigma}\bm{n}$ is the viscous shear stress along the substrate, where 
$\bm{P}_\text{t} = \bm{1} - \bm{n} \otimes \bm{n}$ is the projector on the tangential plane normal to $\bm{n}$, thus any pressure term
$-p \bm{1}$ in the stress tensor $\bm{\sigma}$ does not contribute. Finally, $l_s$ is the slip length and for $l_s \rightarrow 0$ one recovers the 
no-slip boundary condition. 

Reference\ \cite{Grawitter24} demonstrates that the viscous shear stresses $\bm{\sigma}\bm{n} + p \bm{1}$ that all act on the droplet vertices
are represented by $\bm{G}_0^{(1)} \dot{\bm{R}}_0$, where we introduced $\bm{G}_0^{(1)}$ in Eq.\ (\ref{eq.friction_droplet}) as part of 
$\bm{G}_0$ in the dynamic equations\ (\ref{eq.dynamic}). To implement 
the second term on the left-hand side
%the right-hand side 
of the Navier condition (\ref{eq.Navier}), 
we need to add a friction matrix $\bm{G}_0^{(2)}$ to $\bm{G}_0$ and define $\bm{G}_\text{slip}$. Thus we have
\begin{equation}
\bm{G}_0^{(2)} =  \bm{P}_{\|0}^\text{t} \bm{Q} \bm{P}_{\|0} 
% \bm{G}_0^{(2)} =  - \bm{P}_{\|0}^\text{t} \bm{Q} \bm{P}_{\|0} 
\enspace \text{and} \enspace
\bm{G}_\text{slip} =  -  \bm{P}_{\|0}^\text{t} \bm{Q} \bm{P}_{\|\text{sh}}
%\bm{G}_\text{slip} =  \bm{P}_{\|0}^\text{t} \bm{Q} \bm{P}_{\|\text{sh}}
\end{equation}
with
\begin{equation}
Q_{ij} =  \frac{\eta}{l_s} A_i \delta_{ij} \, ,
\end{equation}
which only acts on the tangential velocity components at the liquid-sheet interface. The operators $\bm{P}_{\|0}$, $\bm{P}_{\|\text{sh}}$ project
these tangential velocities out of the respective velocity vectors $\dot{\bm{R}}_0$, $\dot{\bm{R}}_\text{sh}$. With these additional
friction matrices one fully satisfies the Navier condition since 
$- \bm{\nabla}_{\bm{R}_0}  (\mathcal{F}_0 + \mathcal{F}_\text{cons})$ in Eq.\ (\ref{eq.dynamic})
does not possess any tangential components besides at the contact line.

While the Navier condition is a kinematic constraint, we also have to balance tangential viscous shear stresses of the fluid with the
in-plane elastic stresses in the sheet, $\bm{P}_\text{t} \bm{\sigma}\bm{n} = - \delta \mathcal{F}_\text{d}/\delta\bm{r}_\text{sh}$.
So the Navier condition becomes $- \delta \mathcal{F}_\text{d}/\delta\bm{r}_\text{sh} =
\hs{-}
 \eta/l_s \bm{P}_\text{t} (\bm{v} - \bm{v}_\text{sh})$
and it is readily satisfied in Eq.\ (\ref{eq.dynamic}) by adding
\begin{equation}
\bm{G}_\text{sh}^{(2)} = 
%- 
\bm{P}_{\|\text{sh}}^\text{t} \bm{Q} \bm{P}_{\|\text{sh}} 
\end{equation}
to the elastic-sheet friction matrix $\bm{G}_\text{sh}^{(1)}$ of Eq.\ (\ref{eq.sheet_friction}).

\subsubsection{Total friction matrix}
\label{subsubsection.total}

%\hs{HS: \textbf{Salik}, I think $\bm{Q}$ should appear with a minus sign in $\bm{G}_0$ and $\bm{G}_\text{sh}$ in contrast to what 
%we write in Josua's paper.} \ga{If so, wouldn't Eq.1 give you something like $-Q\dot{R} +Q\dot{R}_{sh} + \text{stuff} = \text{driving force}$? Then the left hand side is 
%written as $-Q(\dot{R} -\dot{R}_{sh}) + \text{stuff}$. But $-Q(\dot{R} -\dot{R}_{sh})$ is the friction force and that has to be equal to minus the driving force so that 
%$\text{driving force} + \text{friction force}=0$. Am I making a mistake?} \\
%\hs{HS: So what is the point now?}  \\
Collecting all the contributions, we arrive at the following friction matrices that fill the total friction block matrix $\bm{G}$ in 
Eq.\ (\ref{eq.dynamic}). For the full droplet friction matrix $\bm{G}_0 = \bm{G}_0^{(1)} + \bm{G}_0^{(2)}$ we have
\begin{equation}
\bm{G}_0 = \bm{X}^{-1}(\bm{ C}_d + \bm{Y}) +  \bm{P}_\text{cl}^\text{t}\bm{M} \bm{P}_\text{cl} 
%- 
+ \bm{P}_{\|0}^\text{t} \bm{Q} \bm{P}_{\|0} \, .
 \label{eq.friction_droplet_full}
\end{equation}
The full elastic-sheet friction matrix reads
\begin{equation}
\bm{G}_\text{sh} = \bm{G}_\text{sh}^{(1)} + \bm{G}_\text{sh}^{(2)} = \xi_\text{sh} \bm{1} 
% - 
+ \bm{P}_{\|\text{sh}}^\text{t} \bm{Q} \bm{P}_{\|\text{sh}}  \, .
\label{eq.sheet_friction_full}
\end{equation}
and the liquid-sheet friction matrix is
\begin{equation}
\bm{G}_\text{slip} = 
- \bm{P}_{\|0}^\text{t} \bm{Q} \bm{P}_{\|\text{sh}} \, .
\end{equation}

\subsection{Final dynamic equations}
\label{subsec.final_dynamics}

For a complete solution of the dynamic equations (\ref{eq.dynamic}), we still need to determine the
Lagrange parameters $\lambda_i$ introduced in Section\ \ref{subsec.cons} and formalized in Eq.\ (\ref{eq.Fcons2}).
We use here a method, which we employed in Ref.\ \cite{Grawitter21} to implement the volume constraint following Ref.\ \cite{Alinovi18},
and generalize it to all our constraints. The idea is to move $- \bm{\nabla}_{\bm{R}}  \mathcal{F}_\text{cons}$ in Eq.\ (\ref{eq.dynamic})
to the left-hand side. For this we enhance the velocity vector $\dot{\bm{R}}$ by the column vector $\bm{\Lambda} = 
[\lambda_0, \lambda_1, \ldots, \lambda_N]$ and introduce the constraint matrix 
$\bm{G}_\text{cons} = [\bm{\nabla}_{\bm{R}} g_0(\bm{R}), \bm{\nabla}_{\bm{R}} g_1(\bm{R}), \ldots, \bm{\nabla}_{\bm{R}} g_N(\bm{R})]$.
Then, the enhanced system of dynamic equations reads,
\begin{equation}
\left[ \begin{array}{cc}
\bm{G} & \bm{G}_\text{cons} \\[.5ex]
\bm{G}_\text{cons}^\text{t}& \bm{0}
\end{array}
\right]
\left[  \begin{array}{c} 
\dot{\bm{R}} \\ [.5ex]
\bm{\Lambda}
\end{array}
\right]
=
\left[  \begin{array}{c} 
- \bm{\nabla}_{\bm{R}}  (\mathcal{F}_0 + \mathcal{F}_\text{sh}) \\[.5ex] 
\bm{0}
\end{array}
\right] \, .
\label{eq.dynamic_enhanced}
\end{equation}
While the upper row represents the original dynamic equations, the lower row states that for each constraint
 $\bm{\nabla}_{\bm{R}} g_i(\bm{R}) \cdot \dot{\bm{R}} = 0$, which indeed needs to be fulfilled. Thus, by inverting the enhanced friction 
 matrix, we are able to calculate $\dot{\bm{R}}$ and thereby the dynamics of the combined droplet-sheet system.
%\\ \hs{HS: Do we need to be more specific about the vectors $\bm{\nabla}_{\bm{R}} g_i(\bm{R})$ ?}

We need to discuss a final point. The droplet and the sheet have their own meshes. However, for implementing the connectivity constraint at
the droplet-sheet interface, the droplet and sheet vertices should coincide. To achieve this, we start with the position vectors
$\bm{r}_{\text{sh}j}^{(I)}$ of the sheet vertices and determine the elastic forces acting on them, $\bm{\nabla}_{\bm{r}_{\text{sh}j}^{(I)}}
\mathcal{F}_\text{sh}$. Then, we perform a weighted interpolation to determine from these forces, the elastic force at the position $\bm{r}_{j}$
of a droplet vertex. The weighted interpolation is performed with the %program Surrogates.jl from Julia.
Julia program package Surrogates.jl. It uses the Inverse Distance Surrogate
method and weighs the forces from sheet vertices surrounding $\bm{r}_{j}$ according to their inverse distance\ga{.}
%\ga{GA:Do we need to go so deep in the details? Can't we simply state that we perform a weighted interpolation?}. 
After solving the dynamic  equations (\ref{eq.dynamic_enhanced}), the sheet velocities $\dot{\bm{r}}_{j}$ are interpolated back on the 
velocities $\dot{\bm{r}}_{\text{sh}j}^{(I)}$ of the sheet vertices, which then perform a time step.
%\ga{are then marched forward in time.}

Finally, the dynamic equations (\ref{eq.dynamic_enhanced}) are discretized on a mesh of gridpoints, which we introduce in the following
subsection. The discretization of the elastic in-plane deformation and bending forces are explained in 
Section\ \ref{subsubsec.sheet_energy}.

%
%\hs{- Comment about discretization:  \hs{?????}  Points below are already included}\\
%Regarding the sheet, the Skalak forces are $-\nabla_{\boldsymbol{R}{\text{sh}}}\mathcal{F}d$ are computed analytically per triangle 
%from the deformation gradient $\boldsymbol{F}$ and the undeformed lengths and angles from each face. The bending forces are 
%computed analytically per edge from the dihedral angle $\theta{ij}$ following the thesis of Kr\"uger (cite kruger2012). The IDW interpolation 
%is built per geometry update and used for both sheet-droplet force transfer as well as for the velocity coupling.
%

\subsection{Non-dimensionalization, pa\-ra\-me\-ters, and simulation details}
\label{subsec.para}

\textbf{Non-dimensionalization}
We non-di\-men\-sionalize the dynamic equations (\ref{eq.dynamic}) as in Ref.\ \cite{Grawitter24}. We rescale lengths by 
the radius $R_0$ of the contact line on an initially plane substrate and energies by $\gamma_\text{lg} R_0^2$. To rescale time, we choose
$\tau_0=9\gamma_\mathrm{lg}^{-1}\eta R_0\ln(\zeta/l_\mathrm{s})$, which results from the Cox-Voinov law for the velocity of the contact
line and is an estimate for the time needed to move a distance $R_0$. Used in Eq.\ (\ref{eq.dynamic}), the droplet properties
only depend on the equilibrium contact angle $\theta_\text{eq}$ as well as the respective ratios $l_s/R_0$ and $\zeta/R_0$ for the slip 
length and the phenomenological length $\zeta$. The non-dimensional shear viscosity becomes $\tilde \eta = [9\ln(\zeta/l_\mathrm{s})]^{-1}$.

\textbf{Droplet parameters}
For the droplet parameters we follow Refs.\ \cite{Grawitter21,Grawitter24} and base them on a 90\%-glycerol/10\%-water mixture
for which de Ruijter\ \emph{et\ al.}~\cite{deruijter97} measured in experiments  $\ln(\zeta/l_\mathrm{s})=44$, 
$\eta=209\,\mathrm{mPa}\,\mathrm{s}$, $l_\mathrm{s}=1\,\mathrm{nm}$, $\gamma_\mathrm{lg}=65.3\,\mathrm{mN}\,\mathrm{m}^{-1}$,
and mass density $\rho=1.24\,\mathrm{g}\,\mathrm{ml}^{-1}$. In particular, to rescale the elastic constants of the sheet, we need 
$\gamma_\mathrm{lg}$ and the length scale $R_0$, for which we set $R_0 = 100 \mu\text{m}$.

\textbf{Elastic-sheet parameters}
The elastic moduli introduced in Section\ \ref{subsubsec.sheet_energy} for the thin elastic sheet can be linked to the elastic properties 
of the bulk elastic material. Typically, these properties are described by the Young modulus $E$ that quantifies the relative elongation due to 
a tensile stress, $\sigma = E \Delta L/L$, and Poisson's ratio that quantifies the relative lateral contraction, 
$\nu = - (\Delta d/ d) / (\Delta L/L )$. Es an example, we take here an SIS (Styrol-Isopren-Styrol) elastomer used in Ref.\ \cite{Schulman15} 
with $E = 1 \text{MPa}$ and $\nu= 0.49$, %thus 
the latter indicating that the bulk material is nearly incompressible \cite{Lee19}. 
Further\-more, we can calculate the 
elastocapillary length, introduced in Section\ \ref{subsubsec.droplet_free} as $l_\text{el} = \gamma_\text{lg} / E = 0.06 \mathrm{\mu m}$.
%\\ \hs{HS: Where are these values actually from? I don't find them so explicitely in the PRL of Schulmann.}

\begin{figure} %[ht!]
  \centering
    \includegraphics[width= 1\columnwidth]{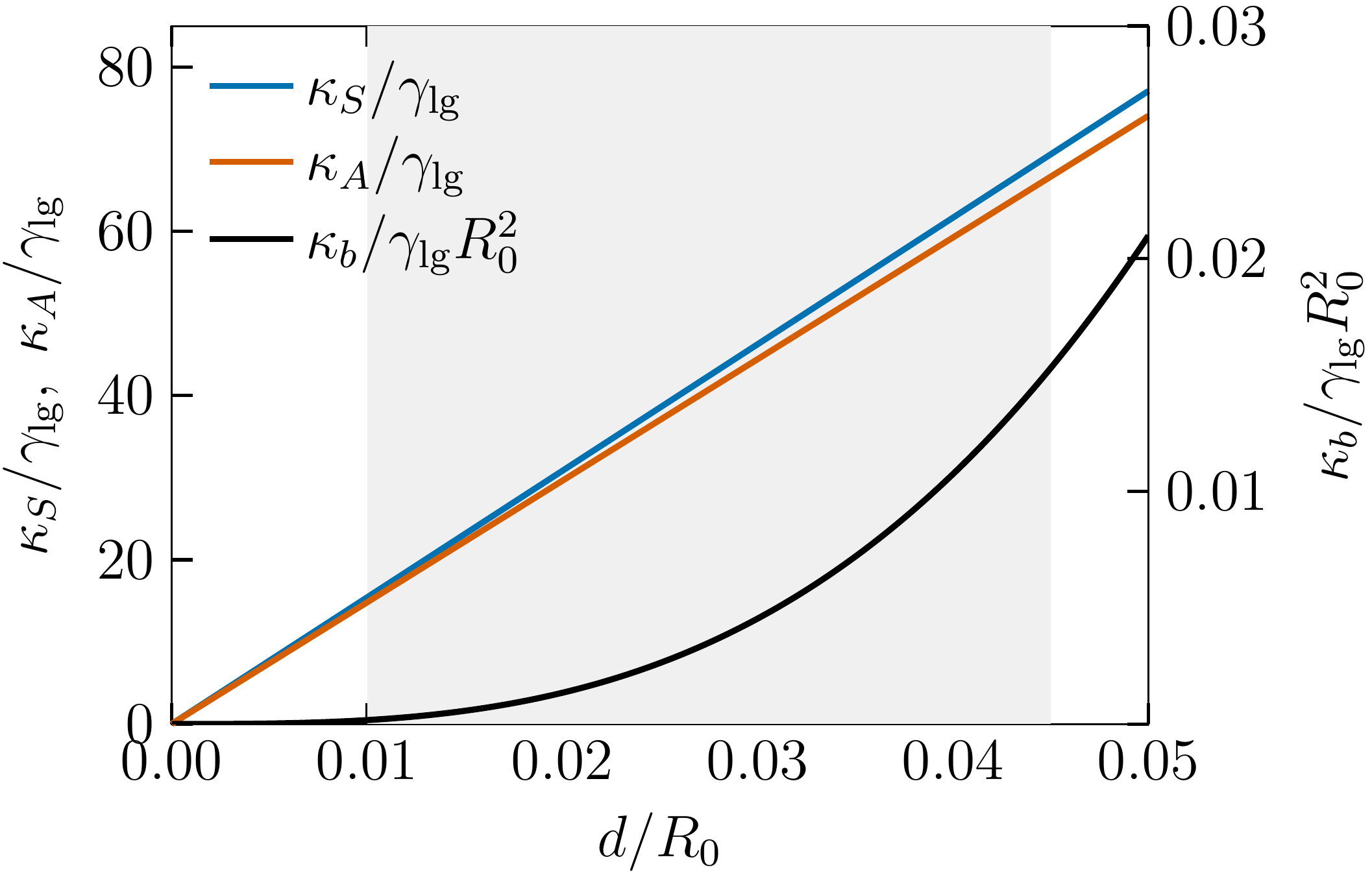}
\caption{The shear ($\kappa_S$), area ($\kappa_A$), and bending ($\kappa_b$) elastic moduli plotted \emph{vs.} film thickness $d$. 
The gray area indicates the thickness range used in our work.
%\gaa{What is the cream/beige region?}
}
\label{fig.elastic_moduli}
\end{figure}

Shear, area, and bending moduli of a thin sheet of thickness $d$ can be related to the bulk properties. Typically, one derives formulas
within linear elasticity theory\ \cite{LandauLifshitzElasticity}.
%\\ \hs{HS: Where are the formulas from?} \\
Matching them to the moduli of the Skalak and Helfrich free energies, we have
\begin{equation}
\kappa_S = \frac{3}{2} \frac{Ed}{1+\nu} \, , \,\,
\kappa_A = \frac{3}{2} \frac{\nu E d}{1-\nu^2} \,\, \text{and} \,\,
\kappa_b = \frac{1}{12} \frac{ E d^3}{1-\nu^2}
\end{equation}
In Fig.\ \ref{fig.elastic_moduli} we plot all three elastic moduli \emph{versus} sheet thickness for the chosen parameters.
We vary the sheet thickness in the range $d = 1 - 4.5 \mathrm{\mu m}$, which is indicated in the plot.
%\\ \hs{HS: %1. Salik, I would need the figure. Can you please make it.
%2. \textbf{Salik}, are the formulas correct. Please check.} \\
While the two in-plane elastic moduli grow linearly in the sheet thickness, the bending modulus increases with $d^3$. So at small thicknesses,
it becomes easier to bend the sheet compared to shearing it.
In fact, at the end of Section\ \ref{subsec.clamped} we will comment that the bending energy is always negligible compared to the
in-plane deformation energy, in particular, for our largest thickness. This is also stated in literature\ \cite{Davidovitch18}.
%\\ \hs{HS: In Sect. 3.1, we have the full numbers}

%\ga{GA: I don't understand this bit. Why do we mention Section 3.1 if we already say it here?}
%
%
%\\ \hs{HS: Stress absolute smallness, even for cirvature radius $R=R_0$ and deformations of 1\% \cite{Davidovitch18}.}

%\hs{HS: \textbf{Salik}, what is the value for the friction coefficient $\xi_\text{sh}$ in Eq. (13)?}
%\\

Finally, we set the friction coefficient of the sheet in Eq.\ (\ref{eq.sheet_friction}) to $\xi_{\text{sh}}=10^{-3} \, \gamma_{\text{lg}}\tau_0/R_0^2$,
which results in a relaxation time much shorter than the characteristic time scale $\tau_0$ of droplet relaxation. Thus, the sheet performs
a quasi-static dynamics.

%\hs{HS: \textbf{Salik}, what is the value for the line tension $\tau$ in Eq. (2). }
%\\

%\hs{HS: \textbf{Salik}, any parameter missing?}
%\\

\textbf{Initial condition} Regardless which equilibrium contact angle $\theta_\text{eq}$ we set, we always start the simulations with a droplet 
sitting on the plane
%\ga{a planar} 
sheet. 
The droplet begins with the shape of a spherical cap,  a contact angle of $90^\circ$, and %a radius $R_0$ of the contact line,
a contact line with radius $R_0$, which we use as a characteristic length to rescale all other lengths. 
This length is 
%not arbitrary, as it 
fixed by the droplet volume.
% of the droplet $V$ as $R_0 = [2V/(3\pi)]^{1/3}$.} 
Furthermore, the initially plane 
%\ga{planar} 
sheet is %quadratic
a square with a side length $4R_0$ and the 
droplet %is placed in the center.
sitting in its center.

In our simulations we considered different cases. We either clamped all four edges of the sheet or moved them 
apart to achieve a relative extension $\epsilon$ (isotropic stress). 
We also apply a uniaxial stress with different conditions at the lateral edges.
When we implement the relative extension $\epsilon$, we let the droplet first relax during
%\ga{for a} 
time $t=0.05$, which we identified as the typical relaxation time of the 
droplet towards the equilibrium configuration. Then, we pull the sheet edges apart always with the same speed until the relative extension 
$\epsilon$ is reached. This ensures that the droplet reaches the equilibrium state 
and does not exhibit numerical pinning.

%. \ga{GA:as opposed to what? Should we say "correct equilibrium state"? Should we say it avoids stuff like numerical pinning or so?}

%
%We can also mention our strain rate for stretching
%the simulations. We first pre-relax the simulation
%for t = 0.05 which is the time it takes for a clamped
%droplet to relax and then begin stretching. All
%simulations are pulled with the same Òboundary
%speedÓ despite the epsilon we use. Perhaps this is
%best added in the section with the isotropic tension
%simulations.
%

\textbf{Meshes}
All simulations are performed using triangle meshes, which are generated with the three-dimensional finite-element mesh generator gmsh.
We employ coarser meshes for sheet and droplet
%\ga{coarser meshes} 
in Section\ \ref{subsec.clamped} and finer meshes with a resolution increased by a factor four in Sections\ \ref{subsec.isotropic} 
and \ref{subsec.uniaxial}. The coarse mesh representing the droplet contains 307 vertices, 608 triangles, and the mean nearest-neighbour 
distance is $0.157 R_0$. The finer mesh is obtained by one refinement step in the mesh generator. It is made of 1219 vertices, 
2432 triangles and the mean nearest-neighbour distance  is $0.080 R_0$. 
%\ga{GA: does it make sense to talk about the refinement step? Unless we somehow need to specify that we do multiple later 
%on or so, I would simply state the properties of the mesh. Otherwise, we need to define what the refinement step is exactly.}
The coarse mesh %of
representing the sheet contains 513 vertices and 944 triangles, while the fine mesh is made of 1969 vertices and 3776 triangles.
%\\ \hs{HS: \textbf{Salik} Can you check the nearest-neighbor distance, please. For the coarser mesh it seems too high.}

%\hs{HS: \textbf{Salik}, What are our mesh properties? What is the number of vertices or triangles (??) for droplet and elastic sheet?
%\\ Here is the old text of Josua.} \\
%All simulations are performed using a surface mesh of 1199~vertices which, in their initial configuration, have an average distance of
%$0.094\,R_0$ from the nearest neighbor.
%
%\hs{HS:} standard mesh, and finer mesh with resolution increased by a factor four

\section{Results}
\label{sec.results}

%\hs{HS: Introductory sentences.
%Test cases for our theoretical approach.}

We use the extended BEM to determine
%We are interested in 
the equilibrium shape of a droplet that sits on an elastic sheet. The shape is determined by a number of control parameters and can
be tuned via various protocols for the boundary conditions of the sheet.
% To disentangle these effects, 
In Section\ \ref{subsec.clamped}
%Wwe begin by characterizing the equilibrium shape for a simple protocol, where the edges of the elastic sheet are clamped. We 
we clamped all edges of the elastic sheet and show
%observe 
how morphological metrics such as the height and the total macroscopic contact angle depend on intrinsic
properties of the sheet, 
%that are intrinsic to the sheet, 
in particular, its thickness. In Section\ \ref{subsec.isotropic}
%we then demonstrate how the same sheet may result in differing equilibrium droplet shapes when using a more complex protocol 
%whereby the sheet is stretched. 
we isotropically stretch the sheet and determine the radial profiles of azimuthal and radial stresses.
Finally, in Section\ \ref{subsec.uniaxial} we stretch the sheet along only one direction using different boundary conditions
for the lateral edges and characterize the elongated shape of the droplet.

%We further observe different phenomenology when the sheet is stretched in only one direction, rather than both of them.

\subsection{Clamped elastic sheet}
\label{subsec.clamped}

\begin{figure*}%[ht!]
  \centering
    \includegraphics[width= 1 \textwidth]{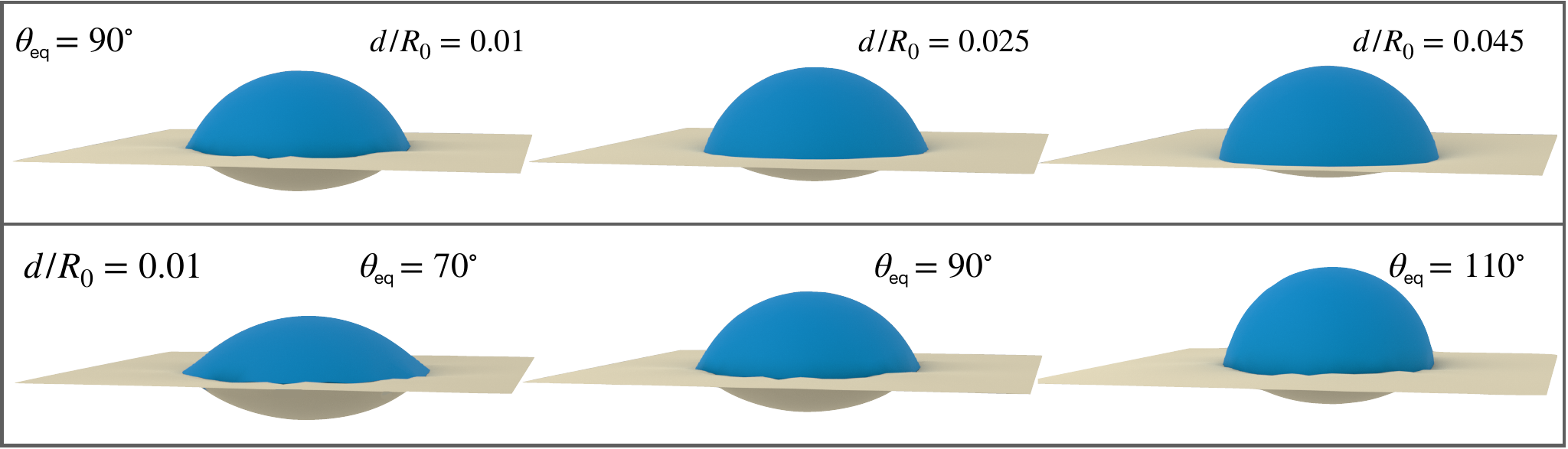}
\caption{Droplets sitting on a thin elastic sheet. Upper row: Equilibrium contact angle $\theta_\text{eq} = 90^\circ$ and different values of sheet thickness $d/R_0$.
Lower row: $d/R_0 = 0.01$ and different values of $\theta_\text{eq}$.
% \ga{For small thickness $d/R_0 = 0.01$, we observe small ripples around the contact line, which we attribute to a non-physical 
%rendering artifact.} \ga{GA:Other parameters missing? Elastic moduli?} 
}
\label{fig.clamped_snapshots}
\end{figure*}

\begin{figure*} %[ht!]
  \centering
    \includegraphics[width= 0.8\textwidth]{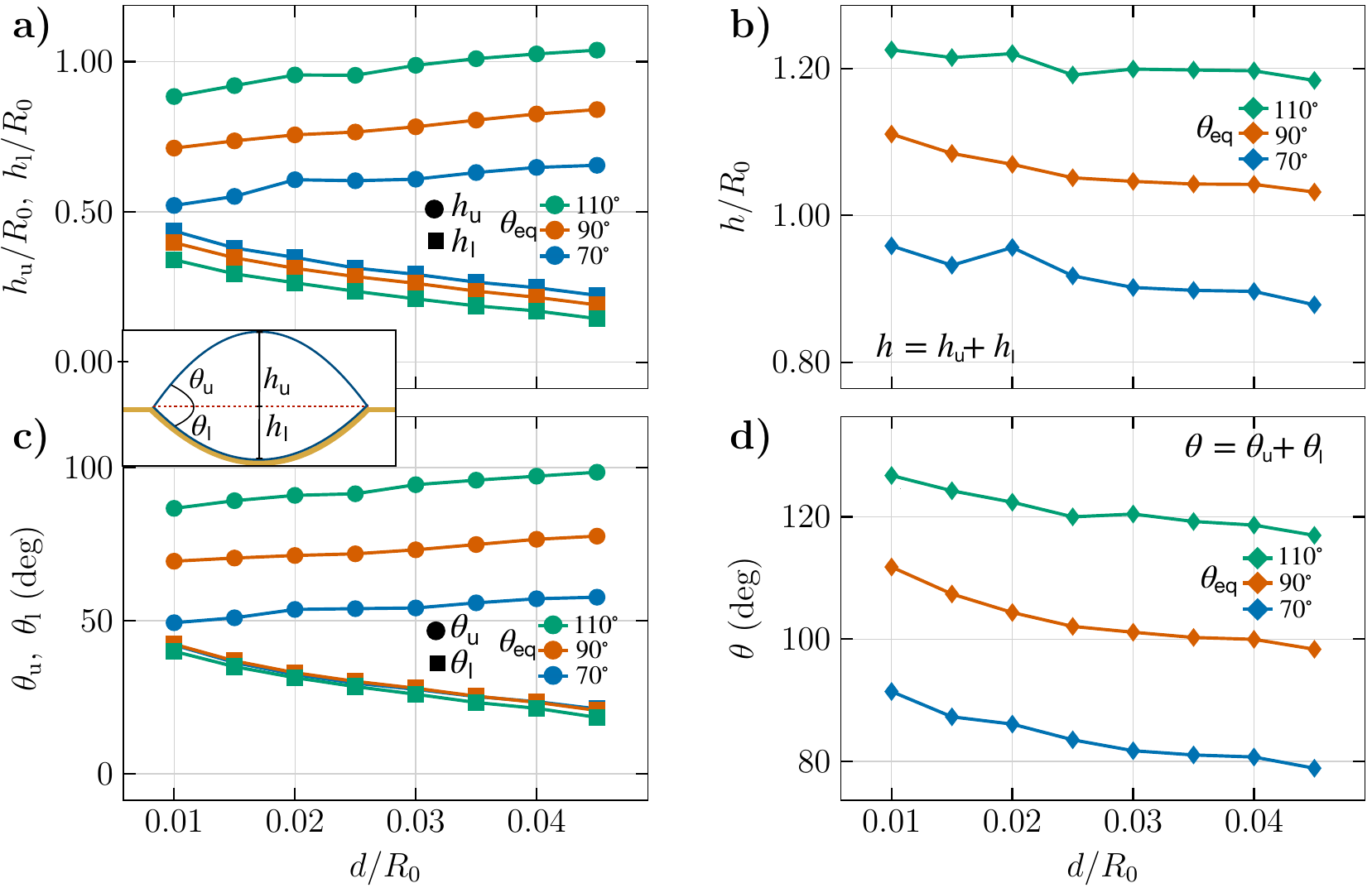}
\caption{%Metrics
Morphological metrics of the droplet sitting on a sheet plotted versus sheet thickness $d/R_0$.
(a) Height of lower ($h_\text{l}$) and upper ($h_\text{u}$) part of the droplet and 
(b) total height $h = h_\text{l} + h_\text{u}$ for several values of $\theta_\text{eq}$.
(c) Macroscopic opening angle of lower ($\theta_\text{l}$) and upper ($\theta_\text{u}$) part and (d) total macroscopic contact angle 
$\theta = \theta_\text{l} + \theta_\text{u}$ for several values of $\theta_\text{eq}$. 
%\ga{GA:Other parameters missing? Elastic moduli?}
}
\label{fig.clamped_metrics}
\end{figure*}

We clamped the four edges of the elastic sheet and %study
studied in detail the resulting shape of the droplet sitting on the sheet for
different equilibrium or Young's angles $\theta_\text{eq}$ and sheet thicknesses $d$. The Laplace pressure in the droplet deforms
the sheet and the droplet assumes a lens shape. The resulting equilibrium configurations in Fig.\ \ref{fig.clamped_snapshots} give 
a first impression of the phenomenology.
%\ga{It is visible in the first row that f}
For increasing sheet thickness (first row), meaning increasing stiffness of the sheet, the droplet sinks less into the sheet and the lens 
shape becomes strongly asymmetric with respect to the horizontal plane defined by the contact line. Tuning the equilibrium angle 
(second row), the droplet sinks in more for smaller values of Young's angle, because the contact area with the sheet is larger.
%\ga{as a result of a larger droplet-sheet} contact \ga{area.} %with the sheet is larger.
%\hs{HS: Revise: Ripples due to rendering. Don't appear in the mesh.\\
For small thickness $d/R_0 = 0.01$, we observe small ripples around the contact line. They do not appear in the mesh of the 
sheet and, therefore, are an artifact when rendering the smooth surface from the mesh.

%from rendering the mesh to generate the surface.
%\\ \hs{HS: We should keep the former sentence here.}

%If we increase the resolution by a factor four,
%the ripples move closer together. They could be a numerical artifact or indicate a wrinkling instability \cite{Huang07}. To decide this question,
%very time-consuming simulations are needed, which we postpone to future work where we plan to study wrinkling patterns around 
%droplets \cite{Huang07}.}
%\\ \hs{HS: Salik is this correct so far.}

In Fig.\ \ref{fig.clamped_metrics} we study the droplet shape in more detail %for varying sheet thicknesses and Young's angles
%using different metrics.
by presenting several morphological metrics, which we plot \emph{versus} sheet thickness $d$ for three values of Young's angle.
% and their variation with the sheet thickness and the Young's angle.}  
The contact line defines a plane, which we use to introduce the heights of the upper and lower part of the droplet 
[see inset between plots (a) and (c)]. Furthermore, assuming spherical caps for the two parts and
fitting circle segments to vertical cross sections of the droplets,
% \ga{that are perpendicular to the contact line}, 
we define upper and lower opening angles [plot (c)].
Their sums
%\ga{with} their sums %
 define
%\ga{defining} 
a macroscopic contact angle \cite{Fortes82}. 
Plot (a) clearly demonstrates the observed phenomenology. For thicker sheets $h_l$ decreases and $h_u$ increases, 
while the total height $h=h_u + h_l$ [plot (b)] does not vary much. This is  interesting, since the droplet transitions from a more symmetric
to a highly non-symmetric shape with increasing thickness (see Fig.\ \ref{fig.clamped_snapshots}).
%
%\ga{GA: Do we want to refer to Fig. (a), (b), etc. separately, or should we always do Fig. 1 (a), 1 (b), etc.?}
%\ga{Plot (a) shows how the morphology of the droplet depends on the thickness of the sheet, with thicker sheets promoting droplets showing 
%lower values of $h_l$ and higher values of $h_u$. This measurement is consistent with a droplet that is increasingly less able to sink into the 
%sheet, owing to the sheet's increasing resistance to deformation (Fig. \ref{fig.elastic_moduli}). Plot (b) shows the total thickness of the 
%droplet $h=h_u + h_l$, which curiously depends very weakly on the sheet thickness, even though the droplet transitions from a highly 
%symmetric shape to a highly non-symmetric shape with increasing thickness (Plot \ref{fig.clamped_snapshots})}
%
A similar trend as for $h_l$ and $h_u$ is observed for $\theta_l$ and $\theta_u${, respectively [plot (c)]. 

Interestingly, the curves of  $\theta_l$ for different $\theta_\text{eq}$ are almost identical,
%run nearly the same,
%\ga{being} 
thus independent of Young's angle, and also the curves of $h_l$ do not vary much with $\theta_\text{eq}$.
A possible reason is that the shape of the lower part of the droplet is strongly determined by the elastic sheet, while
$\theta_\text{eq}$ influences both parts.
Therefore, controlling $\theta_\text{eq}$, provides a method for selectively tuning the shape of the upper droplet half.
%\\ \hs{HS: We do not need to bring here explicit examples how to control $\theta_\text{eq}$.}\\
Thereby, the optical properties of the droplet, used as a liquid lens, can be tuned, as we will discuss in Section\ \ref{subsec.isotropic}.

% \ga{We further remark that $h_l$ varies with $\theta_{eq}$ in a much weaker fashion than $h_u$. Such a decoupling provides an 
%avenue for selectively tuning the curvature of only one of the droplets faces (the one not in contact with the surface), by controlling 
%$\theta_{eq}$ by e.g. changing the composition of the vapor phase, or by changing the wetting properties of the sheet via a light 
%source \cite{something}. Such control enables a more precise tuning of the optical properties of the droplet /liquid lens, as we shall 
%discuss later on. GA: Can you guys include here the appropriate references for the light-activated wettability change?}
%\\ \hs{Reason?} \ga{GA:No idea, actually. The $h_l$ are also very similar. Curious.} \\  

The total macroscopic contact angle in plot (d) is always above $\theta_\text{eq}$.
It does not agree with $\theta_\text{eq}$ due to the lens shape of the droplet, as we discuss in the next paragraph.
% hinting to the fact that the droplet 
%deviates from spherical-cap shape 
%shows a lens shape due to the deformation of the sheet. \ga{GA: I don't get the previous reasoning.} 
%\ga{Interestingly, the total macroscopic contact angle $\theta$ does not equal $\theta_\text{eq}$.} %It should slowly converge 
Nonetheless, it should slowly
%is expected to 
converge towards $\theta_\text{eq}$ for large thicknesses, where the sheet is hardly deformed
% \ga{(nearly planar)}
%. It stays planar 
and the drop
%has 
assumes the typical spherical-cap shape of a planar substrate.
However, this limit is not realized at $d = 0.045 R_0$ as the non-zero values for $h_l$ show.

The deformation of the sheet close to the contact line is nicely illustrated in the inset of Fig.\ \ref{fig.microscopic}. The liquid-gas surface
tension of the free droplet surface pulls at the contact line. The normal component is balanced by the normal bending stress due to the 
negative curvature of the sheet.
%The contact line sits exactly where the curvature of the sheet changes sign. The normal sheet tension is zero, 
The tangential sheet stresses at the contact line cancel each other, and the microscopic contact angle or Young's angle should be realized. 
This is verified in the main plot of Fig.\ \ref{fig.microscopic}, where we plot the measured contact angle \emph{versus} 
Young's angle for sheets with thickness $d/R_0 = 0.02$.  To generate the plot, we increased the linear mesh resolution by
a factor two to better resolve the sheet deformation close to the contact line.

\begin{figure}%[ht!]
  \centering
    \includegraphics[width= 0.95 \columnwidth]{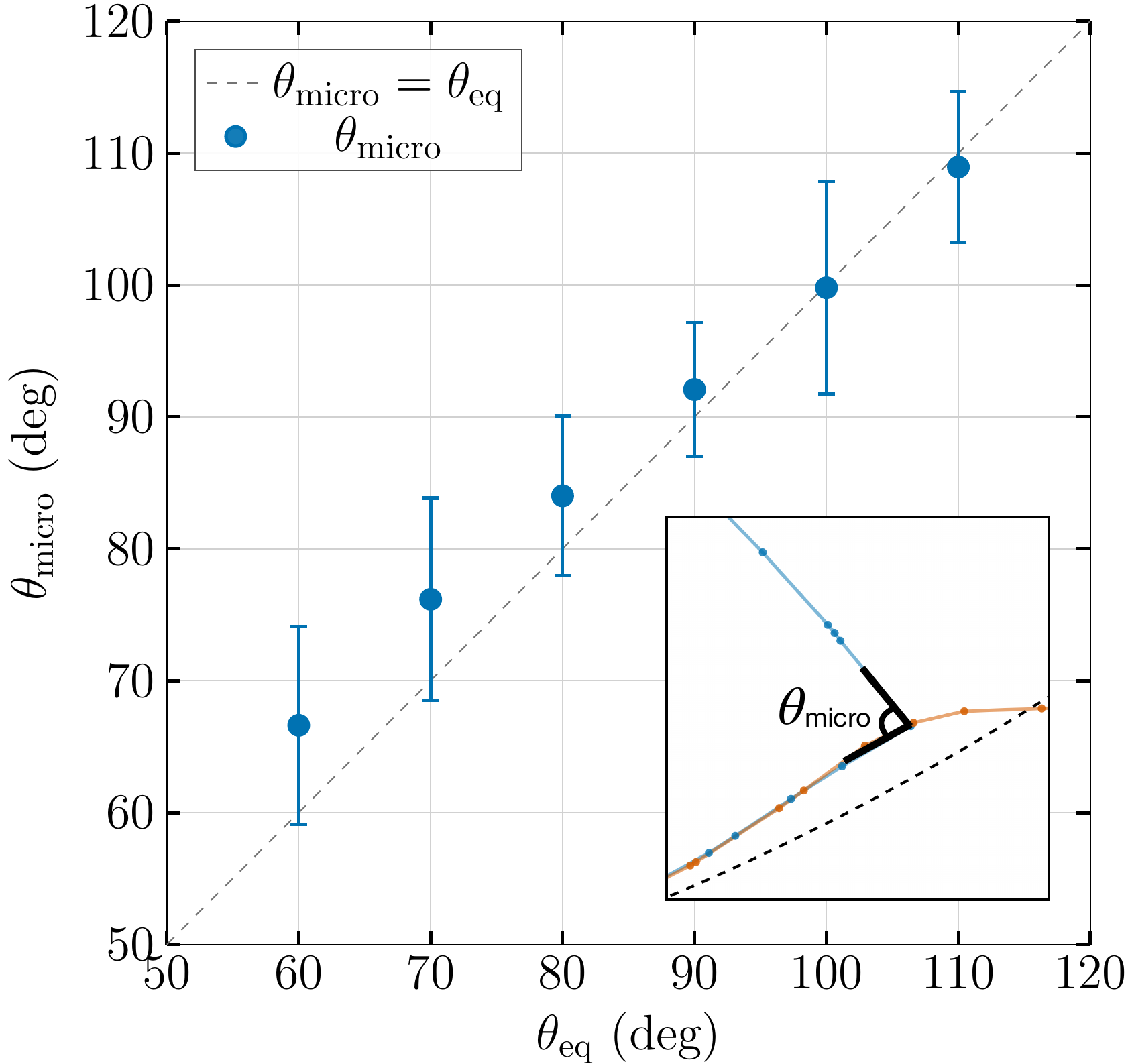}
\caption{Microscopic contact angle plotted \emph{vs} equilibrium contact angle for a sheet of thickness $d=0.02R_0$. The dashed line 
shows $\theta_\text{micro} = \theta_\text{eq}$. Inset: The angle $\theta_\text{micro}$ is determined from local tangents of the 
liquid-gas interface (blue) and sheet (brown) at the contact line. The dashed line shows the circular fit to the droplet-sheet interface 
extended to the region of the contact line.
%\ga{GA:Other parameters missing? Elastic moduli?}
}
\label{fig.microscopic}
\end{figure}

%1. We first discuss different values of thicknesses and equlibrium angles.\\
%2. Show 6 snapshots in a two-column figure: \\
%$\qquad$ first row: $\theta_{eq} = 90$ (or other value) and d = 0.01, 0.025, 0.045\\
%$\qquad$ second row: $d=0.01$ and three $\theta_{eq}$ values.\\
%Mention crumpling ..... Reason: mesh not fine enough or physical?\\
%3. Show the diagrams with the metrix over two-columns (see friday report)\\
%Explain the definition of the quantities, mention how the angle is determined.\\
%The macroscopic contact angles (determined by fitting a circle to the droplet cross section assuming a sperical-cap shape): 
%the total angle deviates strongly from Young's angle. sheet deformation and deviation from sperical cap shape
%close to the contact line

We add two final notes about the elasticity regime. 
First, from Fig.\ \ref{fig.clamped_metrics}(a) we realize that the height $h_l$ of the lower part of the 
droplet assumes values up to $h_l=0.5 R_0$. Assuming that the radius of the circular contact line remains roughly the same, the area of the 
lower spherical cap is $\pi (R_0^2 + h_l^2) = 1.25  \pi R_0^2$. So the increase in area compared to the flat sheet-droplet interface before the 
deformation is  25\%, which corresponds to a linear increase in one direction by 12\%. This is at the upper limit of the validity of the linear 
elasticity regime.
Second, for the same droplet geometry, if we compare bending and an estimate of the area dilatation energy using the moduli in Fig.\ \ref{fig.elastic_moduli} 
at our largest thickness, we realize that bending energy is always negligible compared to the in-plane deformation energy.
 
%Stress that we are beyond the linear elasticity regime: Really?
%area of sphere cap: $A=2\pi R h_\text{l}$, where $R$ is the sphere radius,
%better: $A= \pi (R_0^2 + h^2)$ \\
%$h=0.5 R_0$ and keeping $R_0$ gives: $1.25  \pi R_0^2$.

%estimate for bending energy ....

\subsection{Isotropically stretched sheet}
\label{subsec.isotropic}

\begin{figure*}%[ht!]
  \centering
    \includegraphics[width= 0.9\textwidth]{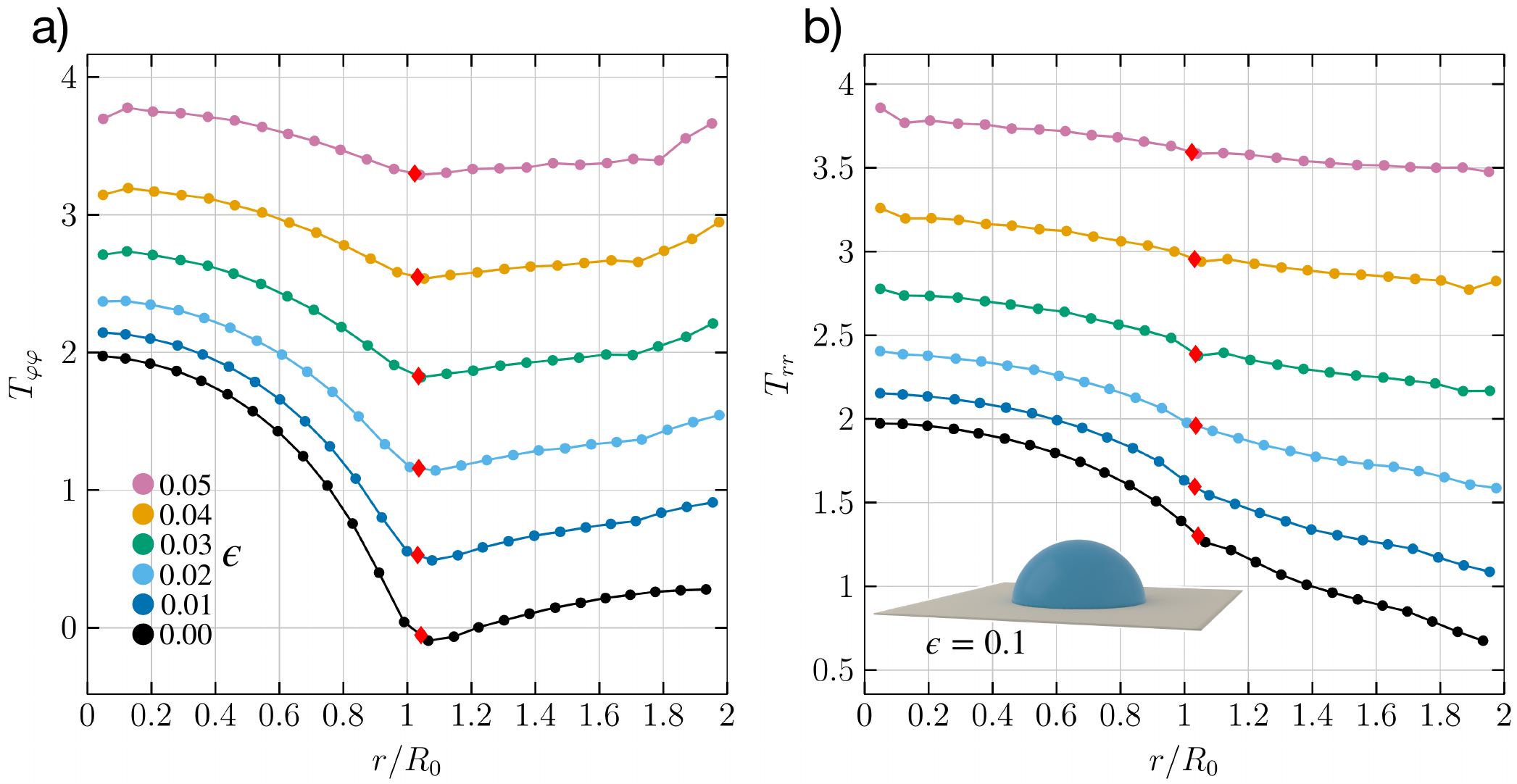}
\caption{Azimuthal (a) and radial (b) stress components plotted \emph{vs.} radial distance from the vertical axis for different
relative extensions $\epsilon$. Here, $T_{\varphi\varphi} = \bm{e}_\varphi \cdot \bm{T} \bm{e}_\varphi$,
%\ga{$\sigma_{\varphi\varphi} = \bm{e}_\varphi \cdot \bm{\sigma} \bm{e}_\varphi$}, 
where $\bm{e}_\varphi$ is 
the azimuthal base vector, and $T_{rr} = \tilde{\bm{e}}_r \cdot \bm{T} \, \tilde{\bm{e}}_r$, where $\tilde{\bm{e}}_r$
%\ga{$\sigma_{rr} = \tilde{\bm{e}}_r \cdot \bm{\sigma} \, \tilde{\bm{e}}_r$}, where $\tilde{\bm{e}}_r$ 
is tangential to the sheet pointing radially outward. The red diamonds indicate the location of the contact line.
Inset in b): At an extension $\epsilon = 0.1$ (stress components are not shown), the sheet is roughly planar. Other parameters are
sheet thickness $d=0.02 R_0$ and $\theta_\text{eq} = 90^\circ$. 
%\\ \hs{HS: Elastic moduli are all introduced in Sect. 2.4.}
%
%\ga{GA:Other parameters missing? Elastic moduli?} \ga{GA: We need to change $T$ to $\sigma$. GA: Was the idea to differentiate 
%between the 3D cauchy stress from the droplet and the 2d cauchy stress in the sheet? Maybe we can do $\sigma$ for the droplet 
%and $\sigma^{\prime}$ for the sheet?}
}
\label{fig.tension}
\end{figure*}

\begin{figure}%[ht!]
  \centering
    \includegraphics[width= 0.95 \columnwidth]{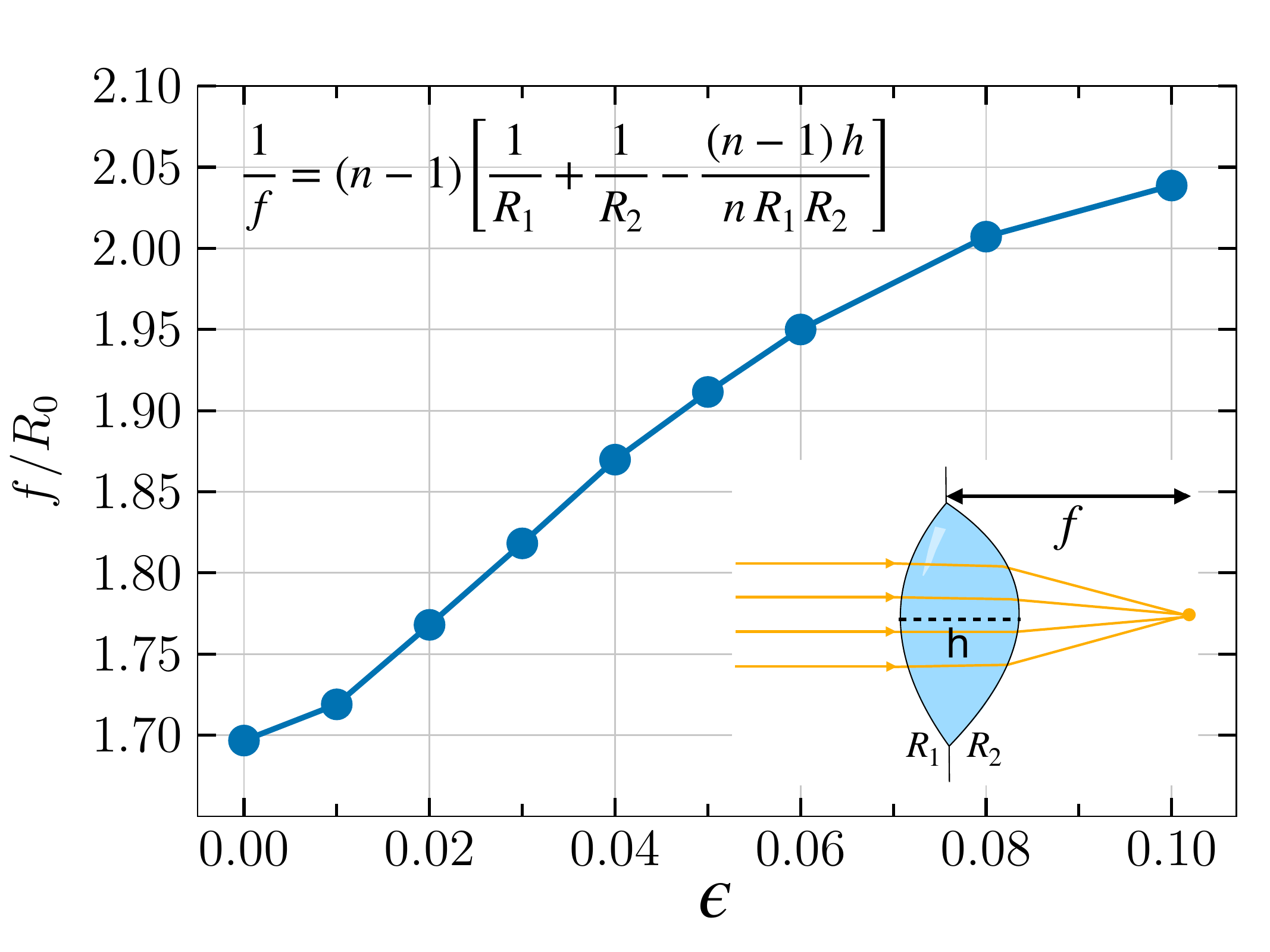}
\caption{Focal length $f$ of the lens-shaped droplet plotted versus the relative extension $\epsilon$.
Lower inset: Illustration of the focal length; $h$ is the thickness of the droplet and $R_1$, $R_2$ are the two curvature 
radii of the two droplet halves. Upper inset: The lens maker's equation of the focal length. The refractive index $n=1.46$ of 
the 90\%-glycerol/10\%-water mixture is used \cite{Rheims97}. 
%\ga{GA:Other parameters, elastic moduli?}
}
\label{fig.focal}
\end{figure}

%\hs{Salik:\\
%We can also mention our strain rate for stretching the simulations. We first pre-relax the simulation for t = 0.05 which is the time it 
%takes for a clamped droplet to relax and then begin stretching. All simulations are pulled with the same ``boundary speed'' despite 
%the epsilon we use. Perhaps this is best added in the section with the isotropic tension simulations.}

We also studied droplets on elastic sheets, which we stretch in both directions up to a defined relative extension $\epsilon$.
Because of the increasing tension within the sheet, the lens-shaped droplet is ``lifted up'' 
%\ga{raised from the sheet} 
until the sheet 
%\ga{latter} 
is flat again and the droplet assumes its equilibrium shape 
%as it were sitting on 
 for a plane substrate. In the inset of Fig.\ \ref{fig.tension}(b) we illustrate this situation for 
$\epsilon = 0.1$ with the equilibrium angle $\theta_\text{eq} = 90^\circ$. 
The attached video  in the supplemental material shows how the droplet first sinks into the sheet, which then is slowly stretched 
up to $\epsilon = 0.1$.
The main plots (a) and (b) in Fig.\ \ref{fig.tension} show 
the relevant stress components as a function of radial distance $r$ from the vertical through the center of the droplet
%\ga{center of the droplet/sheet}
% averaged over the azimuthal angle $\phi$, 
for different relative extensions $\epsilon$ up to $0.05$. To calculate the stress components, we rely on the Cauchy stress tensor 
of the sheet given in spatial (Eulerian) coordinates:
%\begin{equation}
%\bm{T} = \frac{2}{\Delta} \, \bm{F} \, \frac{\partial f_\text{d}}{\partial \bm{C}}  \, %\bm{F}^\text{t} \, .
%\end{equation}
\begin{equation}
\bm{T} = \frac{2}{\Delta} \, \bm{F} \, \frac{\partial f_\text{d}}{\partial \bm{C}}  \, \bm{F}^\text{t} \, .
\end{equation}
%\ga{GA: Same as before wit hthe nomenclature. Plus $\bm{T}$ was already used for the oseen stresses.}
Here, $2 \partial f_\text{d} / \partial \bm{C}$ is the second Piola-Kirchoff stress tensor, which derives from the Skalak free-energy density 
in Eq.\ (\ref{eq.Skalak}) in material coordinates. The deformation gradient tensor $\bm{F}$ transforms it to spatial coordinates and 
$\Delta = \| \bm{F}_1 \times \bm{F}_2 \|$ is the two-dimensional determinant to account for changes in local area in the deformed sheet. 
It is written as cross 
product of vectors $\bm{F}_\alpha$ with components $F_{i\alpha}$ from the deformation gradient tensor. We take the stress tensor $\bm{T}$ 
%\ga{$\bm{\sigma}$} 
and determine the diagonal components along the azimuthal ($T_{\varphi \varphi}$) and radial ($T_{rr}$) directions, where radial 
here means in the plane of the sheet. We plot the stress components, averaged over the azimuthal angle $\varphi$, in Fig.\ \ref{fig.tension}(a) 
and (b), respectively.

In general, the stresses increase with extension $\epsilon$ and the variation along $r$ is largest for smallest $\epsilon$. Here, the 
droplet causes the largest deformation of the sheet. The contact line (red diamonds) is roughly situated close to an inflection point, 
where the curvature in the stress curves changes sign. While for small $\epsilon$ the droplet is clearly recognizable in both stress curves,
the radial component $T_{rr}$ becomes nearly linear in $r$ with increasing $\epsilon$.
We observe that at large $\epsilon$ close to $r=0$ the stress curve is not horizontal as required by cylindrical symmetry, which is due to
the fact that only a small number of stress values are available for averaging. Also, close to $r=2$ the component $T_{\varphi\varphi}$ bends 
upwards more strongly; this is at the edge of the sheet, where the cylindrical symmetry is no longer valid. Finally, for $\epsilon = 0$
there is a small region close to the contact line, where $T_{\varphi\varphi}$ becomes negative. Such compressional stresses are necessary to
observe wrinkling in the sheet \cite{Huang07}. However, in our case these stresses seem to be too small since we do not observe any
wrinkling.

Liquid lenses offer the ability to change their optical properties very rapidly in contrast to conventional lens systems and have already
reached commercial use\ \cite{Fogle20}.
%
% https://www.optotune.com/products/focus-tunable-lenses/
%
Now, our study of lens-shaped droplets enable us to have full access 
%to the lens shape and, therefore, 
to the optical properties of the lens through its shape and, most importantly, to tune the optical properties by applying tension to the
elastic sheet. The focal length (see lower inset of Fig. \ref{fig.focal}) is the most important quantity of a lens since it determines the 
magnification and angle of view. 
%\hs{\cite{....}}. 
We used the lens maker's equation (see upper inset of Fig.\ \ref{fig.focal}) to calculate the 
focal length for the lens-shaped droplet as a function of the relative extension (see Fig.\ \ref{fig.focal}). Applying tension 
%to the elastic sheet 
offers the possibility to tune the focal length, dynamically and reversibly, by about 15\%. Interestingly,
this range agrees roughly with the variation of the main lens in a typical smartphone. 
%\hs{\cite{ ... }}.
%(GA: we also need a citation), despite the two systems using vastly differing lens-shaping mechanisms}. 
The curve in Fig.\ \ref{fig.microscopic} has an S shape and tends towards a constant value for increasing $\epsilon$ that corresponds 
to the spherical-cap shape on a plane substrate.

%\ga{The use of liquid droplets/lens as microfluidic microscopes has seen a rise in popularity, already having seen commercial 
%use \cite{something} GA: SAlik, can you put down the company that you know about? With citation.} Our study of lens-shaped droplets 
%enable us to have full access to the lens shape, and therefore to the optical properties of the lens. Arguably one of the most important 
%is the focal length, defined as the length from the lens at which parallel light rays are focused (see inset of Fig. \ref{fig.focal}). 
%We also calculated 
%\ga{ This quantity can be calculated via the lens-maker equation, which we reproduce in Fig. \ref{fig.focal}, where we also show the 
%variation of the } focal length %for the lens-shaped droplet
%as a function of the relative extension. %(see Fig.\ \ref{fig.focal}).
%The focal length of a lens determines \ga{crucial properties such as} the magnification and angle of view \ga{\cite{something} GA: 
%We need a citation here}. Applying tension to the elastic sheet offers the possibility
%to tune the focal length \ga{dynamically and reversibly} by about 15\%.  \ga{Curiously, this change} agrees roughly with the variation 
%of the main lens in a \ga{typical} smartphone \ga{\cite{something} (GA: we also need a citation), despite the two systems using vastly 
%differing lens-shaping mechanisms}. 
%The curve in Fig.\ \ref{fig.microscopic} has an S shape and tends towards a constant value for increasing $\epsilon$ that corresponds to the 
%spherical-cap shape on a plane substrate.

\subsection{Uniaxially stretched sheet}
\label{subsec.uniaxial}

\begin{figure*} %[ht!]
  \centering
    \includegraphics[width= 0.95\textwidth]{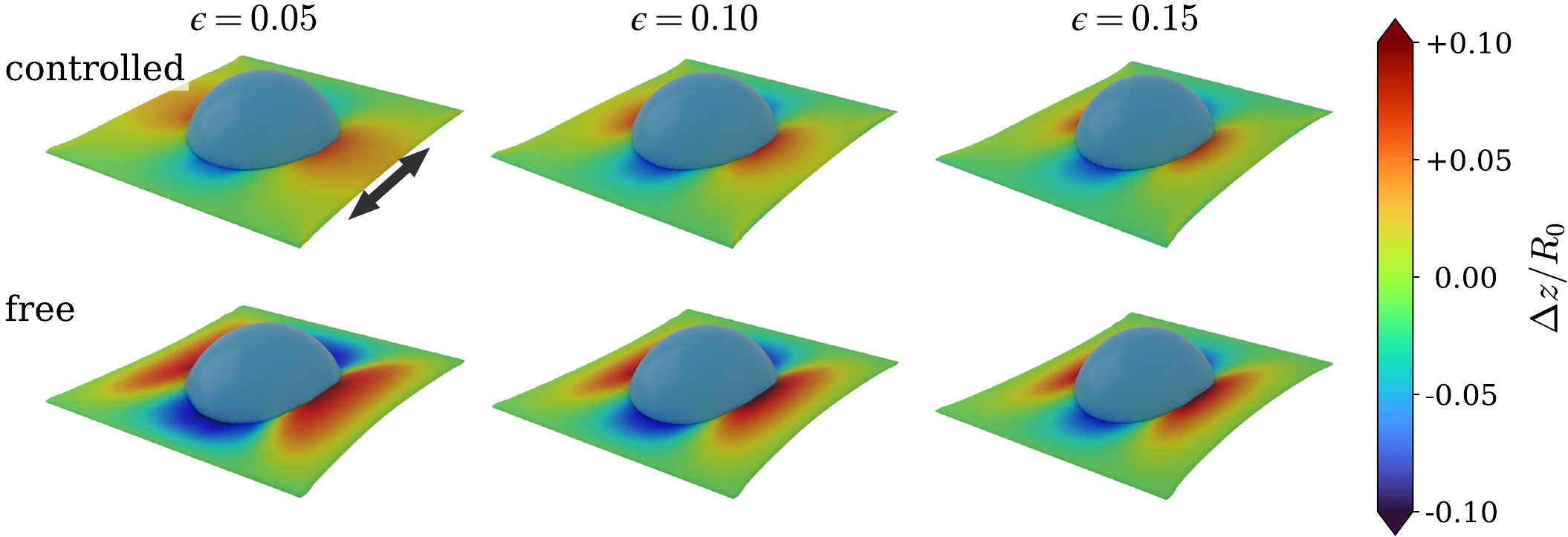}
\caption{%Metrics
Droplets sitting on sheets stretched in one direction, as indicated by the double arrow, for controlled (top) and free (bottom)
boundary conditions of the lateral edges. The columns correspond to three relative extensions $\epsilon$. The heatmap rendering
of the sheet shows the vertical displacement $\Delta z$ relative to the initially plane sheet.
%\\ \hs{HS: Salik, in the upper left sheet, please draw a double arrow next to the right edge, indicating the stretching direction.}
%
%
%Heatmap rendering of steady state simulations for a uniaxially stretched sheet under
%two different boundary conditions on the non-stretched edges: controlled (top row,
%$y$-edges driven at the Poisson contraction rate
%$\varepsilon_y = -\nu_{\mathrm{eff}} \varepsilon_x$ so the sheet remains in
%clean uniaxial stress) and free (bottom row, $y$-edges completely unconstrained).
%Each column corresponds to a different applied stretch
%$\varepsilon \in \{0.05, 0.10, 0.15\}$. Color encodes the sheet's vertical
%displacement $\Delta z$ from the initial flat configuration $(z = 0)$. The free configuration exhibits significant out-of-plane buckling flanking the %droplet. This is because the unconstrained lateral edges allow for Poisson contraction to develop
%into a compressive state under the capillarity of the droplet. The controlled
%configuration suppresses this buckling by prescribing a controlled lateral motion.
%
}
\label{fig.uniaxial}
\end{figure*}

\begin{figure}%[ht!]
  \centering
    \includegraphics[width= 1 \columnwidth]{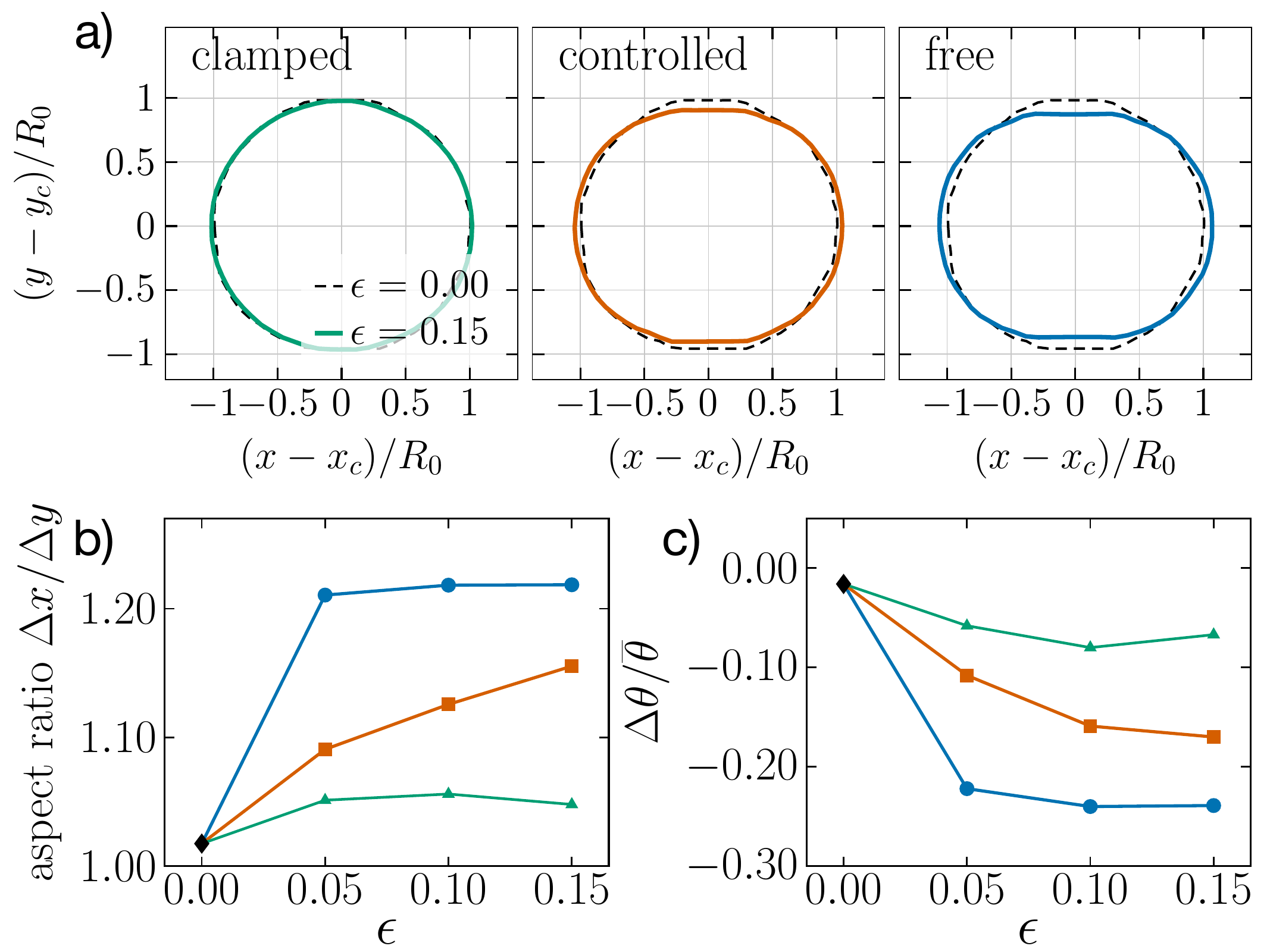}
\caption{(a) Footprints of the contact line of a uniaxially stretched sheet with extension $\epsilon = 0.15$
for the three boundary conditions with clamped, controlled, and free lateral edges.
% for extension $\epsilon = 0.15$. 
The dashed line is the reference case for the clamped sheet with $\epsilon = 0$.
(b) Aspect ratio of the contact line, $\Delta x / \Delta y$,
% ratio of major to minor axis 
and (c) the relative anisotropy of the total macroscopic contact angle, $\Delta \theta / \overline{\theta}$,
plotted as a function of $\epsilon$ for the three boundary conditions. The same color code as in (a) is chosen. 
%\\ \hs{HS: 1. $\epsilon$ is a small letter, can you increase it a bit in the $x$ axis of the plots in (b),(c).
%\\ 2. the words in plot (c) are too small. Perhaps you can skip them completely and make the words larger in the
%footprint in (a). Then, the color code should be clear}
%
%Footprint of the contact line (a), aspect ratio (b), and macroscopic contact angle
%anisotropy (c) illustrating the effects of uniaxial tension with three different
%boundary conditions on the non-stretched edges. Panel (a) shows the steady state contact line at $\varepsilon = 15\%$ (solid line) overlaid 
%on the unstretched reference state at $\varepsilon = 0$ (dashed line) for each of the three boundary conditions (left to right: clamped, 
%controlled, free). The aspect ratio in (b) is
%$\Delta x / \Delta y$, the ratio between the contact line extent in the stretch
%direction and the perpendicular direction. The contact angle anisotropy in (c) is
%defined as
%$\Delta \theta / \bar{\theta}
%= 2(\theta_{x} - \theta_{y})/
%(\theta_{x} + \theta_{y})$,
%where $\theta_{x}$ and $\theta_{y}$ are the total angles
%(upper plus lower contact angle) of the droplet, measured by fitting a circle along $x$-slices and
%$y$-slices. A similar trend follows, droplet anisotropy is strongest 
%\gaa{Can we move the legend for Fig. (a) ($\epsilon$) inside the droplet so that the green curves are not cut-off?}
}
\label{fig.footprint}
\end{figure}

Experiments show that droplets aquire an elongated shape if a uniaxial stress is applied to the sheet\ \cite{Schulman17,Smith21}.
To study this setting, we stretch the elastic sheet along the $x$ direction with a relative extension $\epsilon$, while for the lateral edges we choose
%investigate the outcome of 
three different boundary conditions.
First, we clamp the lateral edges so that they do not move. Second, in what we refer to as the controlled boundary condition,
we let them move inwards by the relative extension $\epsilon_y =  - \nu \epsilon$, where $\nu = 0.49$ is the Poisson ratio of the bulk material. 
%\ga{Note that $\epsilon$ here refers to the relative extension in the $x$ direction.} 
Thus, $\epsilon_y$ is chosen 
%such as %\ga{ that which }%it 
just like it would occur in a uniaxially stretched sheet without a droplet. Third, the lateral edges can move freely. 

For the controlled and free boundary conditions, Fig.\ \ref{fig.uniaxial} shows droplets sitting on uniaxially stretched sheets for three 
values of $\epsilon$. The elongation of the droplets along the stretching direction is clearly visible, in particular
for the case of the free 
lateral edges (bottom). The heatmap on the sheet shows the vertical displacement relative to the plane sheet at the start of the simulation.
Between the free edges and the droplet (bottom row) the sheet folds upward since the edges move inward, while in the stretching direction 
dimples occur. 
This shows that the contact line is not planar; it is elevated along the lateral direction and lowered along the stretching direction. 
Furthermore, the amount of vertical displacement decreases for increasing $\epsilon$, %so 
meaning when stresses are larger. The same qualitative 
behavior can be observed for controlled lateral edges (top row) %
%\ga{albeit with lower values of the vertical displacements when comparing with those of free edges.} 
but the vertical displacements are smaller compared to the free edges.
The latter have larger inward lateral displacements and, therefore, the folds are more pronounced.

In Fig.\ \ref{fig.footprint} we present a quantitative analysis of the elongated droplet for the three boundary conditions. 
For an extension $\epsilon = 0.15$, Fig.\ \ref{fig.footprint}(a) shows the footprint of the contact line, which is formed by projecting the 
contact line on the horizontal plane. 
%for the extension $\epsilon = 0.15$ 
%together with 
The dashed line is the contact line of the unstretched sheet. 
Again, for the free lateral edges (blue, on the right) the elongation of the droplet along the
stretching direction is largest, and for the sheet with clamped lateral edges (green, 
on the left) the elongation is %small
the smallest. In Fig.\ \ref{fig.footprint}(b) the aspect ratio $\Delta x/ \Delta y$ of the major axis to the minor axis of the footprint is 
plotted \emph{versus} $\epsilon$ for all three boundary conditions.
The 
%\ga{maximum observed values of the aspect} 
ratio is close to one for the clamped case (below 1.06) and
% \ga{ $\approx 5\%$}. \ga{Furthermore, we observe this ratio to depend non-monotonically on $\epsilon$, slightly decreasing}  %
even decreases beyond $\varepsilon = 0.1$.
This behavior is expected since for large stresses the sheet becomes planar again, which means the droplet recovers %has 
a circular contact line. Interestingly,
%\ga{T}
the aspect ratio in the case of free lateral edges 
%\ga{takes values that are larger ($\approx 20\%$) but that stay}%
is roughly constant for the 
%\ga{same} 
non-zero values of $\epsilon$
%\ga{. Interestingly, this robustness} occurs 
although the vertical displacements of the sheet do indeed vary.
% \ga{for the same range of $\epsilon$}. 
For the controlled lateral edges the aspect ratio increases with $\epsilon$ throughout the entire range of simulated values of $\epsilon$. 
One major finding of the whole study is that elongation of the droplet along the stretching direction is due to the droplet experiencing 
different sheet stiffnesses along the stretching and the lateral directions, which then determines the elongated shape of the droplet 
when sinking into the sheet. However, for protocols which do clamp the lateral edges, the sheet becomes flat again under large 
extensions and
%However, once the sheet becomes flat again under large extensions and with clamped lateral edges, 
the droplet assumes its spherical-cap shape.
%\ga{From these observations, we conclude that anisotropy (between the stretching and the lateral direction) of the sheet's shape promotes 
%droplet elongation. This anisotropy arises ultimately due to a direction-dependent tension. 
%Protocols which do not promote this anisotropy (such as clamping the lateral edges or stretching the sheet so much that the 
%Laplace pressure of the droplet is negligible) lead to a sheet that takes a flatter shape, and as such induces a droplet shape that 
%increasingly approaches a spherical cap.}

The anisotropy of the droplet shape is also vi\-si\-ble in the total macroscopic contact angles along the stretching ($\theta_x$) and 
lateral ($\theta_y$) directions. 
%extends beyond the contact line. We now show that the total macroscopic contact angles along the stretching ($\theta_x$) 
%and lateral ($\theta_y$) directions may also differ. 
We measure these angles
%We also determined the total macroscopic contact angles along the stretching ($\theta_x$) and lateral ($\theta_y$) directions 
by fitting circles to the upper and lower droplet halves %in both directions, 
as before, but now in both directions. In Fig.\ \ref{fig.footprint}(c) we plot the relative anisotropy
\begin{equation} 
\frac{\Delta \theta}{\overline{\theta}} = \frac{\theta_{x} - \theta_{y}}{(\theta_{x} + \theta_{y})/2}
\end{equation}
\emph{versus} $\epsilon$. The anisotropy is negative indicating that along the stretching direction the droplet shape runs more flat and in 
agreement with the positive aspect ratio of the contact line. The magnitude of the anisotropy is smallest for the clamped case and largest 
for the free lateral edges, where the anisotropy again is nearly constant for the non-zero values of $\epsilon$.
%\backmatter

\section{Conclusions}
\label{sec.concl}

%\gaa{I don't think we should refer to Grawitter24 as a formalized BEM. It implies other BEMs are somehow ad-hoc. 
%Besides, Grawitter24 is still not bullet proof.}
In this article we extended our formalized BEM \cite{Grawitter24}, which we developed earlier to treat the dynamics of droplets on an undulating
substrate. The BEM includes an elastic sheet as a substrate and determines the dynamics of the deforming sheet induced by 
the wetting droplet.
%that is deformed when it is wetted by a droplet.
%to \ga{now} include 
%the deformations of an elastic sheet, 
%when it is wetted by a droplet.
%\gaan elastic substrate in the form of a thin elastic sheet.} 
%Doing so required solving not only the dynamics of the droplet, but also of the deforming sheet.} 
%\ga{The governing equation of our previously-developed BEM, which represents a force balance between minimization of a free energy 
%and friction, was extended }
%This means that besides describing the dynamic 
%droplet configuration, we also needed to include the dynamics of the deforming elastic %sheet in the governing equations. So, in our formalized
%
Thus, in our formalized BEM the total friction matrix is extended by contributions for the elastic sheet and
coupling terms between sheet and droplet that derive from the Navier slip condition and a force balance at the droplet-sheet interface.
%for
%\ga{representing} the \ga{deformations of the} elastic sheet and coupling terms between sheet and droplet. \ga{We %that derive
%derived the later terms} from the Navier slip condition and a force balance at the droplet-sheet interface%.
%\ga{, while the former arises from a modified free energy}. 
The free energy of the total system now includes the Skalak and Helfrich free energies for in-plane and bending deformations of the 
sheet, respectively. Real elastic parameters of an elastomer are chosen to calculate the shear, area, and bending moduli as a function 
of the sheet thickness.

To test our method, we studied three specific 
%cases. 
protocols.
First, we thoroughly characterized the lens shape of the droplet sitting on a clamped
elastic sheet using morphological metrics. They include the heights and macroscopic opening angles of the two droplet halves, which determine
the total height and macroscopic contact angle. Interestingly, %the metrics of the
%the height and opening angle of the 
the response of the lower droplet half %as a function of 
to variations of the sheet thickness %do not vary much with 
is nearly independent of Young's angle. This offers the possibility to selectively tune 
the shape of the upper droplet part while keeping the lower droplet half unchanged. 
Therefore, our finding may open avenues for enhanced control of liquid optical lenses.
%not in contact with the sheet 
%for applications in liquid lenses. 
%\ga{Despite the varying macroscopic contact angle, we found that when z}
Finally, zooming into the region around the contact line, the microscopic contact angle agrees with Young's angle, as expected.

Second, we stretched the sheet in all four directions equally. The droplet is ``lifted up'' 
%\ga{ raised from the sheet}
 until the sheet 
%\ga{latter} 
becomes flat again and assumes
%its typical 
its spherical-cap shape, typical for planar substrates.
Using the stress tensor derived from the Skalak free energy, we are able to extract
%our theory \gaa{wait, what theory? Do you mean theoretical work/simulations?} we are able to derive \gaa{observe? have access to?} 
the radial profiles of azimuthal and radial elastic stresses for several sheet extensions $\epsilon$
from our simulation data.
Droplets on elastic sheets take the form of liquid lenses, which %They 
%such liquid lenses 
are already used in commercial applications. 
The droplet shape, changing with $\epsilon$, manifests itself in an increasing focal length.
%of the liquid lenses. They are already used in com-
%mercials applications and 
Thus, applying tension to the elastic sheet offers the possibility to tune the optical properties of such liquid optical lenses.
%
%Droplets on elastic sheets take the form of a liquid lens, which %They 
%%such liquid lenses 
%are already used in commercial applications. 
%%In this article, we have shown that %and
%Now, applying tension to the elastic sheet offers the possibility to tune 
%%their
%the \ga{lens'} optical properties. \ga{Indeed, we have shown that a change in $\epsilon$ induces a change in the focal length of such a lens.}

 %They are already used in commercials applications.
% applying tension to the elastic sheet offers the possibility to 
%tune %their
%the optical properties. 
From a more fundamental perspective, the droplet creates a deformation field in the elastic sheet. Similar to
capillary interactions, which manifest themselves in the Cheerios effect\ \cite{Vella05}, the sheet will mediate elastocapillary interactions 
between two droplets, which in the simplest setting should be attractive.
%\gaa{Are we sure? The bigger the droplet, the bigger the bulge of the elastic sheet. Maybe it's possible to make a repulsive potential. 
%The cheerios effect is also repulsive if the meniscii of two objects are in opposite direction (meaning one sinks down and the other goes up). 
%Maybe we can get something like that for droplets of different Young's angle. I would refrain from this statement, it seems speculative to me. 
%But maybe it's clear to you.} 
Controlling the tension in the sheet thus offers the possibility of tuning these interactions and, thereby, the assembly of droplets.
% \ga{Applying a uniaxial tension would further result in anisotropic effective interactions and may give rise to novel coarsening dynamics.}

Finally, we stretched the sheet along one direction using different boundary conditions for the lateral edges. Due to the coupling between 
sheet and droplet, the shape of the droplet becomes elongated, the vertical sheet displacement shows upward folds and dimples,  and the 
contact line is no longer planar. We thoroughly characterized the droplet shape by the aspect ratio of the footprint of the contact line and
the anisotropy in the macroscopic contact angle.

% \ga{As we have mentioned before, droplets are known to engage in durotaxis, which leads to the question of how multiple droplets 
%on an elastic sheet may induce each other to move by a deformation-mediated interaction. The tensions displayed in Fig. \ref{fig.tension} 
%could then be used to predict the sign of this effective interaction parameter.} However, the two components of the stress vary in 
%opposite ways with $r$, and the question remains open whether the presence of a deforming substrate aids or hinders the coalescence 
%of droplets. \ga{GA: Is what I am saying true? Have people done these studies?}

The reported work using the extended BEM opens several exciting avenues for future research.
We plan to address wrinkling patterns around droplets, which we already see when we leave all
edges of the sheet free. In a gradient of tension within the sheet, we expect the droplet to perform durotaxis.
%It is also very appealing
The possibility to study elastocapillary interactions between lens-shaped
 %\ga{between}
 droplets and to develop an effective interaction potential 
%\ga{ which may be coarse-grained to an effective potential} 
is also very appealing. Under a uniaxial tension the effective potential for droplets with an elongated shape will be anisotropic
and might cause novel complex dynamics.
Finally, droplets sitting on an elastic sheet can be used as %tunable 
%\ga{optical instruments:} 
liquid optical lenses. 
%We expect further examples of such applications to come up. 
%\gaa{What do you mean with this last sentence?} 
Owing to their highly and easily tunable focal length, these lenses fill a market niche
% that is 
left open by conventional optical instruments. Thus, our work also opens novel avenues for tuning the focal length 
of liquid lenses by controlling the tension applied to the sheet.

\bmhead{Supplementary information}
Below is the link to one video as electronic supplementary material
showing how the droplet is ``lifted up'' 
%\ga{raises from the sheet} 
under increasing tension.
\hs{Link needs to be inserted.}

\bmhead{Acknowledgements}
We thank Sebastian Aland, Stefan Karpitschka, Dirk Peschka, and Uwe Thiele for helpful discussions.
%\gaa{Nobody comes to mind for me}
This work was funded by the German Research Foundation (DFG) within the priority program SPP 2171 (Grant No.\ 505839720).
%
%\hs{??? OLD} We thank Till Welker, Josua Grawitter, Swarnajit Chatterjee, and Fernando Peruani for helpful discussions. 
%We acknowledge financial support from TU Berlin and 
%the German Research Foundation (DFG) as a part of research grant number 462445093. 
We acknowledge support by the Open Access Publication Fund of TU Berlin.

\section*{Declarations}
\subsection*{Funding}
Open Access funding enabled and organized by Projekt DEAL.

\subsection*{Competing interests}
There are no conflicts of interest to declare.

\subsection*{Data availability}
The datasets generated and/or analyzed during the current study are available 
%from the corresponding author 
on reasonable request.

\subsection*{Author contribution}
All authors contributed to the study conception and design. Data collection and analysis were performed by SS. All authors contributed to 
the writing of the manuscript. All authors read and approved the final manuscript.
%\\ \hs{HS: Josua, shall we change this? Please let me know what you prefer.}

%\begin{appendices}
%
%\section{.....}
%\label{sec:app_syssize}

%We now consider what happens for an indeterminate value of $\mathcal{T}_A/\mathcal{T}_{C}$. In this case we can write the total torque as:
%\begin{equation}
%\label{eq:}
%\mathcal{T}_{ij}=\rho R^2\sin{(\phi_i)}\Big(\sin{(\alpha)}\mathcal{T}_C-\alpha \mathcal{T}_A\Big).
%\end{equation}
%Once again, the particle feels no torque when $\phi_0=\pi n$, with $n$ being either $0$ or $1$. However, in cases where $\mathcal{T}_A<\mathcal{T}_C$, it is possible that for a certain value of $\alpha$, $\mathcal{T}_{ij}=0$ regardless of $\phi_i$. Furthermore, the stability of the fixed points is no longer independent of $\alpha$ and instead depends on the values of $\alpha$, $\mathcal{T}_A$, and $\mathcal{T}_C$.

%\end{appendices}


\bibliography{references}% common bib file

@article{Agudo21,
	author = {Agudo-Canalejo, Jaime and Schultz, Sebastian W. and Chino, Haruka and Migliano, Simona M. and Saito, Chieko and Koyama-Honda, Ikuko and Stenmark, Harald and Brech, Andreas and May, Alexander I. and Mizushima, Noboru and Knorr, Roland L.},
	da = {2021/03/01},
	doi = {10.1038/s41586-020-2992-3},
	id = {Agudo-Canalejo2021},
	isbn = {1476-4687},
	journal = {Nature},
	number = {7848},
	pages = {142--146},
	title = {Wetting regulates autophagy of phase-separated compartments and the cytosol},
	ty = {JOUR},
	url = {https://doi.org/10.1038/s41586-020-2992-3},
	volume = {591},
	year = {2021}
}

@article{Aland21,
    author = {Aland, Sebastian and Mokbel, Dominic},
    title = {A unified numerical model for wetting of soft substrates},
    journal = {Int. J. Numer. Methods Eng.},
    volume = {122},
    number = {4},
    pages = {903-918},
    doi = {10.1002/nme.6567},
    url = {https://onlinelibrary.wiley.com/doi/abs/10.1002/nme.6567},
    year = {2021}
}

@article{Alert19,
    author = {Alert, Ricard and Casademunt, Jaume},
    title = {Role of Substrate Stiffness in Tissue Spreading: Wetting Transition and Tissue Durotaxis},
    journal = {Langmuir},
    volume = {35},
    number = {23},
    pages = {7571-7577},
    year = {2019},
    doi = {10.1021/acs.langmuir.8b02037}
}

@article{Alinovi18,
    author = {Edoardo Alinovi and Alessandro Bottaro},
    title = {A boundary element method for Stokes flows with interfaces},
    journal = {J.\ Comp.\ Phys.},
    volume = {356},
    pages = {261-281},
    year = {2018},
    issn = {0021-9991},
    doi = {10.1016/j.jcp.2017.12.004}
}

@article{Andreotti16,
    author = {Andreotti, Bruno and Snoeijer, Jacco H.},
    title = {Soft wetting and the {S}huttleworth effect, at the crossroads between 
             thermodynamics and mechanics},
    journal = {EPL},
    volume = {113},
    number = {},
    pages = {66001},
    year = {2016},
    doi = {10.1209/0295-5075/113/66001}
}

@article{Andreotti20,
    author = {Andreotti, Bruno and Snoeijer, Jacco H.},
    title = {Statics and Dynamics of Soft Wetting},
    journal = {Annu.\ Rev.\ Fluid\ Mech.},
    volume = {52},
    number = {1},
    pages = {285-308},
    year = {2020},
    doi = {10.1146/annurev-fluid-010719-060147}
}

@article{Berge00,
	author = {B. Berge and J. Peseux},
	title = {Variable focal lens controlled by an external voltage:  An application of electrowetting},
	DOI= {epje/v3/p159(e9028)},
	journal = {Eur. Phys. J. E},
	year = 2000,
	volume = 3,
	number = 2,
	pages = {159-163},
}

@article{Bico18,
    author = {Bico, JosÃ© and Reyssat, {\' E}tienne and Roman, BenoÃ®t},
    title = {Elastocapillarity: When Surface Tension Deforms Elastic Solids},
    journal = {Annu. Rev. Fluid Mech.},
    volume = {50},
    number = {1},
    pages = {629-659},
    year = {2018},
    doi = {10.1146/annurev-fluid-122316-050130}
}

@article{Brinker20,
    author = {Manuel Brinker  and Guido Dittrich  and Claudia Richert  and Pirmin Lakner  and Tobias Krekeler  and Thomas F. Keller  and Norbert Huber  and Patrick Huber },
    title = {Giant electrochemical actuation in a nanoporous silicon-polypyrrole hybrid material},
    journal = {Sci. Adv.},
    volume = {6},
    number = {40},
    pages = {eaba1483},
    year = {2020},
    doi = {10.1126/sciadv.aba1483},
    url = {https://www.science.org/doi/abs/10.1126/sciadv.aba1483}
}

@article{Brinker21,
    author = {Brinker, Manuel and Huber, Patrick},
    title = {Wafer-Scale Electroactive Nanoporous Silicon: Large and Fully Reversible Electrochemo-Mechanical Actuation in Aqueous Electrolytes},
    journal = {Adv. Mater.},
    pages = {2105923},
    doi = {10.1002/adma.202105923},
    url = {https://onlinelibrary.wiley.com/doi/abs/10.1002/adma.202105923},
    year = {2021}
}

@article{Chen18,
    title = {Static and dynamic wetting of soft substrates},
    journal = {Curr. Opin. Colloid Interface Sci.},
    volume = {36},
    pages = {46-57},
    year = {2018},
    issn = {1359-0294},
    doi = {10.1016/j.cocis.2017.12.001},
    url = {https://www.sciencedirect.com/science/article/pii/S1359029417301371},
    author = {Longquan Chen and Elmar Bonaccurso and Tatiana Gambaryan-Roisman and Victor Starov and Nektaria Koursari and Yapu Zhao},
}

@ARTICLE{Chen21,
AUTHOR={Chen, Leihao and Ghilardi, Michele and Busfield, James J. C. and Carpi, Federico},
TITLE={Electrically Tunable Lenses: A Review},
JOURNAL={Frontiers in Robotics and AI},
VOLUME={8},
PAGES={166},
YEAR={2021},
URL={https://www.frontiersin.org/article/10.3389/frobt.2021.678046},       
DOI={10.3389/frobt.2021.678046},      
ISSN={2296-9144},   
}

@Article{Davidovitch18,
    author ="Davidovitch, Benny and Vella, Dominic",
    title  ="Partial wetting of thin solid sheets under tension",
    journal  ="Soft Matter",
    year  ="2018",
    volume  ="14",
    issue  ="24",
    pages  ="4913-4934",
    publisher  ="The Royal Society of Chemistry",
    doi  ="10.1039/C8SM00323H",
    url  ="http://dx.doi.org/10.1039/C8SM00323H"
 }

@article{deruijter97,
    title = {Contact Angle Relaxation during the Spreading of Partially Wetting Drops},
    author = {de Ruijter, M. J. and De Coninck, J. and Blake, T. D. and Clarke, A. and
Rankin, A.},
    journal = {Langmuir},
    year = {1997},
    volume = {13},
    pages = {7293},
    doi = {10.1021/la970825v}
}

@incollection{Duprat16,
    author ="Duprat, Camille and Stone, Howard A.",
    title  ="Chapter 6: {E}lastocapillarity",
    booktitle  ="Fluid-Structure Interactions in Low-Reynolds-Number Flows",
    year  ="2016",
    pages  ="193-246",
    publisher  ="The Royal Society of Chemistry",
    address = {Cambridge},
    isbn  ="978-1-84973-813-2",
    doi  ="10.1039/9781782628491-00193"
}

@article{Fortes82,
    author = { M. Amaral Fortes},
    title = {Microscopic and Macroscopic Contact Angles},
    journal = {J. Chem. Soc., Faraday Trans. I},
    volume = {78},
    number = { },
    pages = {101-107},
    year = {1982},
    doi = {10.1039/F19827800101}
}

@article{Gomez20,
    author={Gomez, Hector
    and Velay-Lizancos, Mirian},
    title={Thin-film model of droplet durotaxis},
    journal={Eur. Phys. J. Special Top.},
    year={2020},
    month={Feb},
    day={01},
    volume={229},
    number={2},
    pages={265-273},
    issn={1951-6401},
    doi={10.1140/epjst/e2019-900127-x}
}

@article{Gompper96,
    author = {G. Gompper and D.~M. Kroll},
    title = {Random Surface Discretizations and the Renormalization of the
Bending Rigidity},
    journal = {J.\ Phys.\ I},
    volume = {6},
    number = {10},
    pages = {1305â1320},
    year = {1996},
    doi = {10.1051/jp1:1996246}
}

@article{Gorcum18,
    title = {Dynamic Solid Surface Tension Causes Droplet Pinning and Depinning},
    author = {M. van Gorcum and B. Andreotti and J. H. Snoeijer and S. Karpitschka},
    journal = {Phys. Rev. Lett.},
    volume = {121},
    pages = {208003},
    year = {2018},
    doi = {10.1103/PhysRevLett.121.208003}
}

@article{Grawitter21,
    title = {Steering droplets on substrates using moving steps in wettability},
    author = {Josua Grawitter and Holger Stark},
    journal = {Soft Matter},
    volume = {17},
    pages = {2454},
    year = {2021},
    doi = {10.1039/d0sm02082f},
    doi+an = {=openaccess},
}

@article{Grawitter21b,
    title = {Droplets on substrates with oscillating wettability},
    author = {Josua Grawitter and Holger Stark},
    journal = {Soft Matter},
    volume = {17},
    pages = {9469},
    year = {2021},
    doi = {10.1039/d1sm01113h},
    doi+an = {=openaccess},
}

@article{Grawitter24,
    title = {Steering droplets on substrates with plane-wave wettability patterns and deformations},
    author = {Josua Grawitter and Holger Stark},
    journal = {Soft Matter},
    volume = {20},
    pages = {3161},
    year = {2024},
    doi = {10.1039/d4sm00213j},
    doi+an = {=openaccess},
}

@article{Helfrich73,
    title = {Elastic properties of lipid bilayers. Theory and possible experiments},
    author = {W. Helfrich},
    journal = {Z.\ Naturforsch.\ C},
    volume = {28},
    pages = {693--703},
    year = {1973},
    doi = {10.1515/znc-1973-11-1209}
}

@article{Huang07,
    author = {Jiangshui Huang and Megan Juszkiewicz and Wim H. de Jeu and Enrique Cerda and Todd Emrick and
                    Narayanan Menon and Thomas P. Russell},
     title = {Capillary Wrinkling of Floating Thin Polymer Films},        
     journal = {Science},
     volume = {317},
    number = {5838},
    pages = {650-653},
    year = {2007},
    doi = {10.1126/science.1144616}
}

@article{Kantor87,
    author = {Y. Kantor and D.~R. Nelson},
    title = {Crumpling transition in polymerized membranes},
    journal = {Phys.\ Rev.\ Lett.},
    volume = {58},
    number = {26},
    pages = {2774â2777},
    year = {1987},
    doi = {doi.org/10.1103/PhysRevLett.58.2774}
}

@book{Kim05,
    title = {Microhydrodynamics},
    author = {Sangtae Kim and Sepp J. Karrila},
    year = {2005},
    isbn = {0486442195},
    publisher = {Dover Publications},
    address = {Mineola/NY}
}

@article{Kozyreff23,
  title = {Effect of external tension on the wetting of an elastic sheet},
  author = {Kozyreff, Gregory and Davidovitch, Benny and Prasath, S. Ganga and Palumbo, Guillaume and Brau, Fabian},
  journal = {Phys. Rev. E},
  volume = {107},
  issue = {3},
  pages = {035101},
  numpages = {17},
  year = {2023},
  month = {Mar},
  publisher = {American Physical Society},
  doi = {10.1103/PhysRevE.107.035101},
  url = {https://link.aps.org/doi/10.1103/PhysRevE.107.035101}
}

@book{Krueger12,
    title = {Computer Simulation Study of Collective Phenomena in Dense Suspensions 
             of Red Blood Cells under Shear},
    author = {Tim Kr\"uger},
    year = {2012},
    publisher = {Springer Spektrum, Wiesbaden 2012},
    doi = {10.1007/978-3-8348-2376-2}
}

@article{Kumar18,
    author = {Deepak Kumar  and Joseph D. Paulsen  and Thomas P. Russell  and Narayanan Menon},
    title = {Wrapping with a splash: High-speed encapsulation with ultrathin sheets},
    journal = {Science},
    volume = {359},
    number = {6377},
    pages = {775-778},
    year = {2018},
    doi = {10.1126/science.aao1290},
    url = {https://www.science.org/doi/abs/10.1126/science.aao1290},
}

@article{Kusumaatmaja21,
    author = {Kusumaatmaja, Halim and May, Alexander I. and Knorr, Roland L.},
    title = {Intracellular wetting mediates contacts between liquid compartments and membrane-bound organelles},
    journal = {Journal of Cell Biology},
    volume = {220},
    number = {10},
    pages = {e202103175},
    year = {2021},
    month = {08},
    issn = {0021-9525},
    doi = {10.1083/jcb.202103175},
    url = {https://doi.org/10.1083/jcb.202103175}
}

@book{LandauLifshitzElasticity,
    author = {Landau, L. D. and Lifshitz, E. M.},
    title = {Theory of Elasticity},
    series = {Course of Theoretical Physics, Vol. 7},
    edition = {3rd},
    publisher = {Elsevier Butterworth-Heinemann},
    address = {Oxford},
    year = {1986},
    doi = {10.1016/B978-0-08-057069-3.50003-6}
}

@article{Lee18,
    author = {Lee, I-Ning and Dobre, Oana and Richards, David and Ballestrem, Christoph and Curran, Judith M. and Hunt, John A. and Richardson, Stephen M. and Swift, Joe and Wong, Lu Shin},
    title = {Photoresponsive Hydrogels with Photoswitchable Mechanical Properties Allow Time-Resolved Analysis of Cellular Responses to Matrix Stiffening},
    journal = {ACS Appl. Mater. Interfaces},
    volume = {10},
    number = {9},
    pages = {7765-7776},
    year = {2018},
    doi = {10.1021/acsami.7b18302}
}

@article{Lee19,
    author = {Young-Joo Lee and Seung-Min Lim and Seol-Min Yi and Jeong-Ho Lee and Sung-gyu Kang and
    Gwang-Mook Choi and Heung Nam Han and  Jeong-Yun Sun and In-Suk Choi and Young-Chang Joo},
    title = {Auxetic elastomers: Mechanically programmable meta-elastomers with an unusual 
               Poisson's ratio overcome the gauge limit of a capacitive type strain sensor},
    journal = {Extreme\ Mech.\ Lett.},
    volume = {31},
    number = { },
    pages = {100516},
    year = {2019},
    doi = {10.1016/j.eml.2019.100516}
}

@article{Li25, 
    title={The motion of a thin drop on an elastic sheet}, 
    volume={1022}, 
    DOI={10.1017/jfm.2025.10779}, 
    journal={J. Fluid Mech.}, 
    author={Li, Zhixuan and Ren, Weiqing}, 
    year={2025}, 
    pages={A4}
}

@article{Moffatt64,
    title = {Viscous and resistive eddies near a sharp corner},
    author = {H.~K. Moffatt},
    journal = {J.\ Fluid\ Mech.},
    volume = {18},
    pages = {1-18},
    year = {1964},
    doi = {10.1017/S0022112064000015}
}

@article{Mokbel24,
    author = {Mokbel, Marcel and Mokbel, Dominic and Liese, Susanne and Weber, Christoph and Aland, Sebastian},
    title = {A Simulation Method for the Wetting Dynamics of Liquid Droplets on Deformable Membranes},
    journal = {SIAM\ J.\ Sc.\ Comput},
    volume = {46},
    number = {6},
    pages = {B806-B829},
    year = {2024},
    doi = {10.1137/24M1641142},
    URL = {https://doi.org/10.1137/24M1641142}
}

@article{mueller04,
    title = {Elastic effects on surface physics},
    journal = {Surf. Sci. Rep.},
    volume = {54},
    number = {5},
    pages = {157-258},
    year = {2004},
    issn = {0167-5729},
    doi = {10.1016/j.surfrep.2004.05.001},
    url = {https://www.sciencedirect.com/science/article/pii/S0167572904000408},
    author = {Pierre M\"uller and Andr\'es Sa\'ul}
}

@article{Nair23, 
    title={Equilibrium shapes of liquid drops on pre-stretched nonlinear elastic membranes}, 
    volume={961}, DOI={10.1017/jfm.2023.223}, 
    journal={J.\ Fluid Mech.}, 
    author={Nair, Vineet and Sharma, Ishan and Shankar, V.}, 
    year={2023}, 
    pages={A28}
}

@article{Nguyen10,
    author = {Nguyen,Nam-Trung },
    title = {Micro-optofluidic Lenses: A review},
    journal = {Biomicrofluidics},
    volume = {4},
    number = {3},
    pages = {031501},
    year = {2010},
    doi = {10.1063/1.3460392}
}

@INPROCEEDINGS{Fogle20,
  author={Fogle, Faisal and Cierny, Ondrej and do Vale Pereira, Paula and Kammerer, William and Cahoy, Kerri},
  booktitle={2020 IEEE Aerospace Conference}, 
  title={Miniature Optical Steerable Antenna for Intersatellite Communications Liquid Lens Characterization}, 
  year={2020},
  volume={},
  number={},
  pages={1-13},
  doi={10.1109/AERO47225.2020.9172448}
  }

@article{Park14,
    author = {Park, Su Ji
    and Weon, Byung Mook
    and Lee, Ji San
    and Lee, Junho
    and Kim, Jinkyung
    and Je, Jung Ho},
    title = {Visualization of asymmetric wetting ridges on soft solids with X-ray microscopy},
    journal = {Nat. Comm.},
    year = {2014},
    month = {Jul},
    day = {10},
    volume = {5},
    number = {1},
    pages = {4369},
    issn = {2041-1723},
    doi = {10.1038/ncomms5369}
}

@article{Patel21,
    author = {Patel, Kuntal and Stark, Holger},
    title  = {A pair of particles in inertial microfluidics: effect of shape, softness, and position},
    journal = {Soft Matter},
    year = {2021},
    volume = {17},
    issue = {18},
    pages = {4804-4817},
    publisher = {The Royal Society of Chemistry},
    doi = {10.1039/D1SM00276G}
}

@book{Pozrikidis92,
    author = {C Pozrikidis},
    year = {1992},
    title = {Boundary integral and singularity methods for linearized viscous flow},
    publisher = {Cambridge University Press},
    address = {Cambridge},
    isbn = {0-521-40693-5},
    doi = {}
}

@article{Rheims97,
  title = {Refractive-index measurements in the near-IR using an Abbe refractometer},
  volume = {8},
  ISSN = {1361-6501},
  url = {http://dx.doi.org/10.1088/0957-0233/8/6/003},
  DOI = {10.1088/0957-0233/8/6/003},
  number = {6},
  journal = {Measurement Science and Technology},
  publisher = {IOP Publishing},
  author = {Rheims,  J and K\"{o}ser,  J and Wriedt,  T},
  year = {1997},
  month = {June},
  pages = {601Ð605}
}

@article{Schaaf17,
    author = {Schaaf, Christian and Stark, Holger},
    title = {Inertial migration and axial control of deformable capsules},
    journal = {Soft Matter},
    year = {2017},
    volume = {13},
    issue = {19},
    pages = {3544-3555},
    publisher = {The Royal Society of Chemistry},
    doi  = {10.1039/C7SM00339K}
}

@article{Schroll13,
  title = {Capillary Deformations of Bendable Films},
  author = {Schroll, R. D. and Adda-Bedia, M. and Cerda, E. and Huang, J. and Menon, N. and Russell, T. P. and Toga, K. B. and Vella, D. and Davidovitch, B.},
  journal = {Phys. Rev. Lett.},
  volume = {111},
  issue = {1},
  pages = {014301},
  numpages = {5},
  year = {2013},
  month = {Jul},
  publisher = {American Physical Society},
  doi = {10.1103/PhysRevLett.111.014301},
  url = {https://link.aps.org/doi/10.1103/PhysRevLett.111.014301}
}

@article{Schulman15,
    author = {Schulman, Rafael D. and Dalnoki-Veress, Kari},
    title = {Liquid Droplets on a Highly Deformable Membrane},
    journal = {Phys.\ Rev.\ Lett.},
    volume = {115},
    number = { },
    pages = {206101},
    year = {2015},
    doi = {10.1103/PhysRevLett.115.206101}
}

@article{Schulman17,
    author = {Rafael D. Schulman and  Ren\'e Ledesma-Alonso and Thomas Salez and Elie Rapha\"el and Kari Dalnoki-Veress},
    title = {Liquid Droplets Act as ``Compass Needles'' for the Stresses in a Deformable Membrane},
    journal = {Phys.\ Rev.\ Lett.},
    volume = {118},
    number = { },
    pages = {198002},
    year = {2017},
    doi = {10.1103/PhysRevLett.118.198002}
}

@article{Schulman18,
  title = {Droplets Capped with an Elastic Film Can Be Round, Elliptical, or Nearly Square},
  author = {Schulman, Rafael D. and Dalnoki-Veress, Kari},
  journal = {Phys. Rev. Lett.},
  volume = {121},
  issue = {24},
  pages = {248004},
  numpages = {5},
  year = {2018},
  month = {Dec},
  publisher = {American Physical Society},
  doi = {10.1103/PhysRevLett.121.248004},
  url = {https://link.aps.org/doi/10.1103/PhysRevLett.121.248004}
}

@article{Shanahan85,
    author = {Martin E. R. Shanahan},
    title = {Contact Angle Equilibrium on Thin Elastic Solids},
    journal = {J. Adhesion},
    volume = {18},
    number = {4},
    pages = {247--267},
    year = {1985},
    publisher = {Taylor \& Francis},
    doi = {10.1080/00218468508080461},
    URL = {https://doi.org/10.1080/00218468508080461}
}

@article{Shanahan87,
    author = {Martin E. R. Shanahan},
    title = {Equilibrium of Liquid Drops on Thin Plates; Plate Rigidity and Stability Considerations},
    journal = {J. Adhesion},
    volume = {20},
    number = {4},
    pages = {261--274},
    year = {1987},
    publisher = {Taylor \& Francis},
    doi = {10.1080/00218468708074946},
    URL = {https://doi.org/10.1080/00218468708074946}
}

@article{Shuttleworth50,
    doi = {10.1088/0370-1298/63/5/302},
    year = 1950,
    month = {may},
    publisher = {{IOP} Publishing},
    volume = {63},
    number = {5},
    pages = {444--457},
    author = {R Shuttleworth},
    title = {The Surface Tension of Solids},
    journal = {Proc. Phys. Soc. Sect. A},
}

@article{Skalak73,
    title = {Strain Energy Function of Red Blood Cell Membranes},
    author = {R. Skalak and A. Tozeren and R. P. Zarda and S. Chien},
    journal = {Biophys.\ J.},
    volume = {13},
    issue = {3},
    pages = {245},
    year = {1973},
    doi = {10.1016/S0006-3495(73)85983-1}
}

@article{Smith21,
  title = {Droplets Sit and Slide Anisotropically on Soft, Stretched Substrates},
  author = {Smith-Mannschott, Katrina and Xu, Qin and Heyden, Stefanie and Bain, Nicolas and Snoeijer, Jacco H. and Dufresne, Eric R. and Style, Robert W.},
  journal = {Phys. Rev. Lett.},
  volume = {126},
  issue = {15},
  pages = {158004},
  numpages = {6},
  year = {2021},
  month = {Apr},
  publisher = {American Physical Society},
  doi = {10.1103/PhysRevLett.126.158004},
  url = {https://link.aps.org/doi/10.1103/PhysRevLett.126.158004}
}

@article{Style13,
    author = {Style, Robert W. and Che, Yonglu and Park, Su Ji and Weon, Byung Mook and Je, Jung Ho and Hyland, Callen and German, Guy K. and Power, Michael P. and Wilen, Larry A. and Wettlaufer, John S. and Dufresne, Eric R.},
    title = {Patterning droplets with durotaxis},
    volume = {110},
    number = {31},
    pages = {12541--12544},
    year = {2013},
    doi = {10.1073/pnas.1307122110},
    publisher = {National Academy of Sciences},
    issn = {0027-8424},
    URL = {https://www.pnas.org/content/110/31/12541},
    journal = {Proc. Natl. Acad. Sci. U.S.A.}
}

@article{Toga13,
    author = "Toga, K. Bugra and Huang, Jiangshui and Cunningham, Kevin and Russell, Thomas P. and Menon, Narayanan",
    title = "A drop on a floating sheet: boundary conditions{,} topography and formation of wrinkles",
    journal = "Soft Matter",
    year = "2013",
    volume = "9",
    issue = "34",
    pages = "8289-8296",
    publisher = "The Royal Society of Chemistry",
    doi = "10.1039/C3SM50736J"
}

@article{Vella05,
    author = {Dominic Vella and L. Mahadevan},
    title = {The {``Cheerios effect''}},
    journal = {Am.\ J.\ Phys.},
    volume = {73},
    number = { },
    pages = {817-825},
    year = {2005},
    doi = {10.1119/1.1898523}
}

@article{Wang24,
	author = {Wang, Yanning and Li, Shulin and Mokbel, Marcel and May, Alexander I. and Liang, Zizhen and Zeng, Yonglun and Wang, Weiqi and Zhang, Honghong and Yu, Feifei and Sporbeck, Katharina and Jiang, Liwen and Aland, Sebastian and Agudo-Canalejo, Jaime and Knorr, Roland L. and Fang, Xiaofeng},
	da = {2024/10/01},
	doi = {10.1038/s41586-024-07990-0},
	id = {Wang2024},
	isbn = {1476-4687},
	journal = {Nature},
	number = {8036},
	pages = {1204--1210},
	title = {Biomolecular condensates mediate bending and scission of endosome membranes},
	ty = {JOUR},
	url = {https://doi.org/10.1038/s41586-024-07990-0},
	volume = {634},
	year = {2024}
}

@article{Xu17,
    author = {Xu, Qin
    and Jensen, Katharine E.
    and Boltyanskiy, Rostislav
    and Sarfati, Rapha{\"e}l
    and Style, Robert W.
    and Dufresne, Eric R.},
    title = {Direct measurement of strain-dependent solid surface stress},
    journal = {Nat. Comm.},
    year = {2017},
    month = {Sep},
    day = {15},
    volume = {8},
    number = {1},
    pages = {555},
    issn = {2041-1723},
    doi = {10.1038/s41467-017-00636-y}
}

\end{document}